\documentclass{aa}
\usepackage{graphicx}
\usepackage{natbib,twoopt}
\usepackage[breaklinks=true]{hyperref} 
\usepackage[varg]{txfonts}
 \bibpunct{(}{)}{;}{a}{}{,}    
 \newcommandtwoopt{\citeads}[3][][]{\href{http://adsabs.harvard.edu/abs/#3}%
                                        {\citealp[#1][#2]{#3}}}
 \newcommandtwoopt{\citepads}[3][][]{\href{http://adsabs.harvard.edu/abs/#3}%
                                        {\citep[#1][#2]{#3}}}
 \newcommandtwoopt{\citetads}[3][][]{\href{http://adsabs.harvard.edu/abs/#3}%
                                        {\citet[#1][#2]{#3}}}
 \newcommandtwoopt{\citealtads}[3][][]{\href{http://adsabs.harvard.edu/abs/#3}%
                                        {\citealt[#1][#2]{#3}}}
 \newcommandtwoopt{\citeyearads}[3][][]%
   {\href{http://adsabs.harvard.edu/abs/#3}{\citeyear[#1][#2]{#3}}}
\begin{document}

\title{Tracing large-scale structures in circumstellar disks with ALMA}
\titlerunning{Tracing large-scale structures in circumstellar disks with ALMA}

\author{Jan Philipp Ruge\inst{\ref{inst1},}\thanks{ruge@astrophysik.uni-kiel.de} \and Sebastian Wolf\inst{\ref{inst1}} \and Ana L. Uribe\inst{\ref{inst2},\ref{inst3}} \and Hubert H. Klahr\inst{\ref{inst2}}}
\institute{Universit\"at zu Kiel, Institut f\"ur Theoretische und Astrophysik, Leibnitzstr. 15, 24098 Kiel, Germany \label{inst1} 
\and
Max-Planck-Institut f\"ur Astronomie, K\"onigstuhl 17, 69117 Heidelberg, Germany \label{inst2}
\and
University of Chicago, The Department of Astronomy and Astrophysics, 5640 S. Ellis Ave, Chicago, IL 60637, USA \label{inst3}}

\authorrunning{Ruge et al.}

\date{Received 15. September 2012 / Accepted 16. November 2012 }

 \abstract{Planets are supposed to form in circumstellar disks. The additional gravitational potential of a planet perturbs the disk and leads to characteristic structures, i.e. spiral waves and gaps, in the disk's density profile.}{We perform a large-scale parameter study of the observability of these planet-induced structures in circumstellar disks in the (sub)mm wavelength range for the Atacama Large (Sub)\-Milli\-meter Array (ALMA).}{On the basis of hydrodynamical and magneto-hydrodynamical simulations of star-disk-planet models, we calculated the disk temperature structure and (sub)mm images of these systems. These were used to derive simulated ALMA images. Because appropriate objects are frequent in the Taurus-Auriga region, we focused on a distance of $140 \, \rm pc$ and a declination of $\approx 20^\circ$. The explored range of star-disk-planet configurations consists of six hydrodynamical simulations (including magnetic fields and different planet masses), nine disk sizes with outer radii ranging from $9 \, \rm AU$ to  $225 \, \rm AU$, 15 total disk masses in the range between $2.67 \cdot 10^{-7} \, \rm M_\odot$ and $4.10 \cdot 10^{-2} \, \rm M_\odot$, six different central stars, and two different grain size distributions, resulting in $10{\,}000$ disk models.}{On almost all scales and in particular down to a scale of a few AU, ALMA is able to trace disk structures induced by planet-disk interaction or by the influence of magnetic fields on the wavelength range between $0.4$ and $2.0 \, \rm mm$. In most cases, the optimum angular resolution is limited by the sensitivity of ALMA. However, within the range of typical masses of protoplanetary disks ($0.1\, \rm M_\odot$-- $0.001\, \rm M_\odot$) the disk mass has a minor impact on the observability. It is possible to resolve disks down to $2.67 \cdot 10^{-6} \, \rm M_\odot$ and trace gaps induced by a planet with $\frac{M_\text{p}}{M_\star} = 0.001$ in disks with $2.67 \cdot 10^{-4} \, \rm M_\odot$ with a signal-to-noise ratio greater than three. The central star has a major impact on the observability of gaps, as well as the considered maximum grainsize of the dust in the disk. In general, it is more likely to trace planet-induced gaps in our magnetohydrodynamical disk models, because gaps are wider in the presence of magnetic fields. We also find that zonal flows resulting from magneto-rotational instability (MRI) create gap-like structures in the disk's re-emission radiation, which are observable with ALMA.}{Through the unprecedented resolution and sensitivity of ALMA in the (sub)mm wavelength range, the expected detailed observations of planet-disk interaction and global disk structures will deepen our understanding of the planet formation and disk evolution process.}


\maketitle

\section{Introduction}
\label{sec:intro}

Although an early configuration of the Atacama Large (Sub)\-Milli\-meter Array (ALMA) has become only recently available, its capabilities and its potential for groundbreaking diskoveries have been demonstrated to be enormous \citepads[e.g.][]{2012ApJ...750L..21B}. Already in the ALMA cycle 0 configurations its angular resolution and sensitivity is several times higher than those of the Submillimeter Array. After completion of the entire array, ALMA will allow one to observe the density structure of young circumstellar disks in unprecedented detail.\par 

Planets are supposed to form in these disks \citepads[e.g.][]{2010arXiv1012.5281M}. The two main scenarios are the core accretion model or the nucleated instability \citepads[e.g.][]{2010arXiv1012.5281M} and the planet formation by a gravitational disk instability \citepads[e.g.][]{2007prpl.conf..607D}. Planets orbiting in circumstellar disks interact gravitationally with the surrounding disk gas (\citealtads{1980ApJ...241..425G}; \citealtads{1984ApJ...285..818P}). Planets with planet-to-star mass ratios over $0.0001$ create significant density perturbations in the disks (\citealtads[e.g.][]{2002ApJ...565.1257T}; \citealtads{1997Icar..126..261W} \citealtads{1984ApJ...285..818P}). One of these structures is a spiral density wave on each side of the orbiting planet (e.g. \citealtads{1984ApJ...285..818P}; \citealtads{2006A&A...445..747K}). The inner spiral propagates inwards, while the outer spiral propagates outwards through the disk. In a frame of reference comoving with the planet, these structures are static and move with it. These spirals have low amplitude in density enhancement, typically $<10\%$ of the background disk column density. The other planet-induced structure is a density gap around the orbit of the planet (\citealtads{1980ApJ...241..425G}; \citealtads{1984ApJ...285..818P}). For massive planets, the tidal torques on the surrounding gas overcome the pressure gradient and the viscous spreading, and the gas around the orbital position is pushed away from corotation (\citealtads{1999ApJ...514..344B}; \citealtads{2006Icar..181..587C}; \citealtads{2006MNRAS.370..529D}). The density in the gap can be a very small fraction of the initial density. Even in more evolved systems, like debris disks, giant planets induce characteristic disk structures (see \citealtads{2012A&A...544A..61E} for a detailed study).\par
Grid-based hydrodynamical or smoothed particle hydrodynamical (SPH) simulations (\citealtads{2002A&A...385..647D}; \citealtads{2010A&A...518A..16F}), combined with follow-up radiative transfer simulations (\citealtads{2002ApJ...566L..97W}; \citealtads{2003ApJ...593.1116J}; \citealtads{2005ApJ...619.1114W} \citealtads{2007ApJ...659L.169J}; \citealtads{2009ApJ...700..820J}; \citealtads{2010sf2a.conf...21G}; \citealtads{2012A&A...547A..58G}), have shown that large-scale disk structures are visible through the continuum re-emission and scattered light of the disk. The case of the observability of simulated line emission has been investigated by \citetads{2008ApJ...673L.195S}, \citetads{2010A&A...523A..69R}, and \citetads{2011ApJ...743L...2C}.
For the infrared wavelength range, the observability {of planet-induced, large-scale disk structures} was explored by \citetads{2006ApJ...637L.125V}, \citetads{2005ApJ...619.1114W}, and \citetads{2009ApJ...700..820J}. However, due to the much lower optical depth in the (sub)mm regime, observations in this wavelength range are better suited to directly tracing the density structure of the disk interior. But, for high angular resolution observations in this wavelength range, large interferometer arrays, like ALMA, are required.\par
The intrinsic implementation of the stellar radiation field in the hydrodynamical simulations affects the gap form and enlarges the vertical extension especially of the outer edge of the gap \citepads{2012ApJ...749..153J}. Furthermore, it has been found that in SPH-models gaps are enlarged in comparison to grid-based hydrodynamical simulations \citepads{2010sf2a.conf...21G}. Also, gaps can be enlarged through the detachment of larger dust grains from the gas \citepads{2004A&A...425L...9P}. Therefore, we consider non radiative hydrodynamical simulations as a simpler disk approximation. The additional presence of magnetic fields within the disk influences the planet formation and planet-disk interaction process (e.g. \citealtads{2003MNRAS.339..983P}; \citealtads{2003MNRAS.339..993N}; \citealtads{2004MNRAS.350..829P}; \citealtads{2004MNRAS.350..849N}; \citealtads{2009ApJ...697.1269J}; \citealtads{2012A&A...545A..81P}). In circumstellar accretion disks, it is well known that molecular viscosity is not sufficient for the outward transport of angular momentum and for the transport of mass inwards, as is required for mass to accrete onto the central star (\citealtads{1981ARAuA..19..137P}; \citealtads{1998RvMP...70....1B}). It was shown that hydrodynamical turbulence from a non linear shear-instability cannot transport angular momentum with the required efficiency to support accretion at the observed rates \citepads{2006Natur.444..343J}. Thus, many mechanisms have been proposed to explain the source of the dissipation that drives accretion. Large-scale magnetic fields with open field line configurations can play a role in the vertical transport of angular momentum away from the disk \citepads{2000prpl.conf..759K}. The baroclinic instability has also been proposed as a possible mechanism for angular momentum transport via vortices \citepads{2003ApJ...582..869K}. One of the most promising mechanisms for driving accretion, and certainly the easiest to incorporate in numerical simulations, is the magneto-rotational instability (MRI). Circumstellar Keplerian disks that are permeated by a weak magnetic field are linearly unstable \citepads{1991ApJ...376..214B}. This instability results in the development of turbulence, which can transport mass and angular momentum efficiently \citepads{1998RvMP...70....1B}. A minimum degree of ionization leading to good coupling between the gas and the magnetic field in the disk is required for the MRI to function. In the cold and dusty midplane regions the degree of ionization will be low, possibly producing a dead zone, while in the outer parts of the disk, cosmic rays can penetrate, allowing for the necessary ionization degree (\citealtads{2010ApJ...708..188T}; \citealtads{2010AuA...515A..70D}; \citealtads{2011ApJ...735..122F}), yet here the low densities will make ambipolar diffusion a powerful way to suppress the magneto-rotational instabilities. For the need to incorporate turbulence, we assume that our disks follow the laws of ideal magnetohydrodynamics. More realistic models of non ideal magnetohydrodynamical disk simulations are still under development. \par

\citetads{2002ApJ...566L..97W} show that ALMA is indeed able to trace pre planetary and planet-induced large-scale structures in circumstellar disks and even allows one to observe the circumplanetary accretion region \citepads{2005ApJ...619.1114W}. \citetads{2010MNRAS.407..181C} find that spiral structures in compact self-gravitating disks, which are the same size as the planet-induced large-scale structures, are visible in ALMA observations over a wider range of wavelength. While most of the earlier studies of the observability of planet-induced large-scale structures were mainly focused on exemplary case studies or on the impact of high planetary masses  \citepads[$5\, \rm M_\textup{Jup}$,][]{2012A&A...547A..58G}, we now investigate a large parameter space of stellar, planetary, and disk parameters, to exactly evaluate under which conditions ALMA will allow one to trace characteristic density structures resulting from planet-disk interaction or magneto-rotational instability (MRI) best. Our model setup contains six 3D-hydrodynamical simulations (including magnetic fields and in particular lower planet masses), where gas and dust are homogeneously mixed. It also includes nine disk sizes with outer radii from $9 \, \rm AU$ to  $225 \, \rm AU$, 15 total disk masses in the range between $2.67 \cdot 10^{-7} \, \rm M_\odot$ and $4.10 \cdot 10^{-2}\, \rm M_\odot$, four main sequence and two pre-main-sequence stars and two different grain size distributions. Besides this, 14 ALMA configurations, seven observing wavelengths in the range from $330\, \rm \mu m$ to $3.3\, \rm mm$ and three exposure times are considered.\par
This work is structured as follows. In \S \, \ref{sec:sim} we outline the hydrodynamical and radiative transfer simulation techniques and describe the parameter space of our investigations in detail in \S \, \ref{sec:model}. The data analysis is outlined in \S \, \ref{sec:analysis}. Our results are shown in \S \, \ref{sec:results}. Finally, we present our conclusions in \S \, \ref{sec:conclusion}.

\begin{table*}[t]

\caption{Table of all simulations.}\label{tab:table_sims}
\centering
\footnotesize
    \begin{tabular}{  l  l  l  l  l  l  l  l  l  l  l }
     \hline \hline
    Name & Simulation & $(N_\textup{r},N_{\vartheta},N_{\varphi})$ & $\eta=M_\textup{p}/M_\star$ & $r_\textup{p}$ & r domain & $\vartheta$ domain ($d\vartheta$) & $\varphi$ domain & $h/r$ & SF\\ 
    & & & & [$\rm AU$]  & [$\rm AU$] & [$\rm rad$] & [$\rm rad$] & & \\ 

    \hline

    S1   & HD  & (256,128,256) &     no planet    &  -  & [1,10] & 0.3 & $2\pi$ & 0.07 & unperturbed \\ 
    SP4  & HD  & (256,128,256) & $10^{-4}$ & 5.0 & [1,10] & 0.3 & $2\pi$ & 0.07 & gap    \\ 
    SP1  & HD  & (256,128,256) & $10^{-3}$ & 5.0 & [1,10] & 0.3 & $2\pi$ & 0.07 & gap    \\ 
    \hline
    ST1  & MHD & (256,128,256) &      no planet    &  -  & [1,10] & 0.3 & $2\pi$ & 0.07 & zonal flow \\ 
    SPT4 & MHD & (256,128,256) & $10^{-4}$ & 5.0 & [1,10] & 0.3 & $2\pi$ & 0.07 & p.gap  \\ 
    SPT6 & MHD & (256,128,256) & $10^{-3}$ & 5.0 & [1,10] & 0.3 & $2\pi$ & 0.07 & gap    \\ 

    \hline
     \end{tabular}
 \tablefoot{In the last column, SF refers to "Special Feature" and the terms have the following meaning. p.gap means partial gap, gap means full gap and zonal flow means that there are magnetic field induced density bumps in the disk. The hydrodynamical simulations are characterized as HD and the magnetohydrodynamical ones as MHD. The name gives the link to the density maps shown in Fig. \ref{fig:f1}. }

\end{table*}

\begin{figure*}[!ht]
  \includegraphics[width=8.5cm]{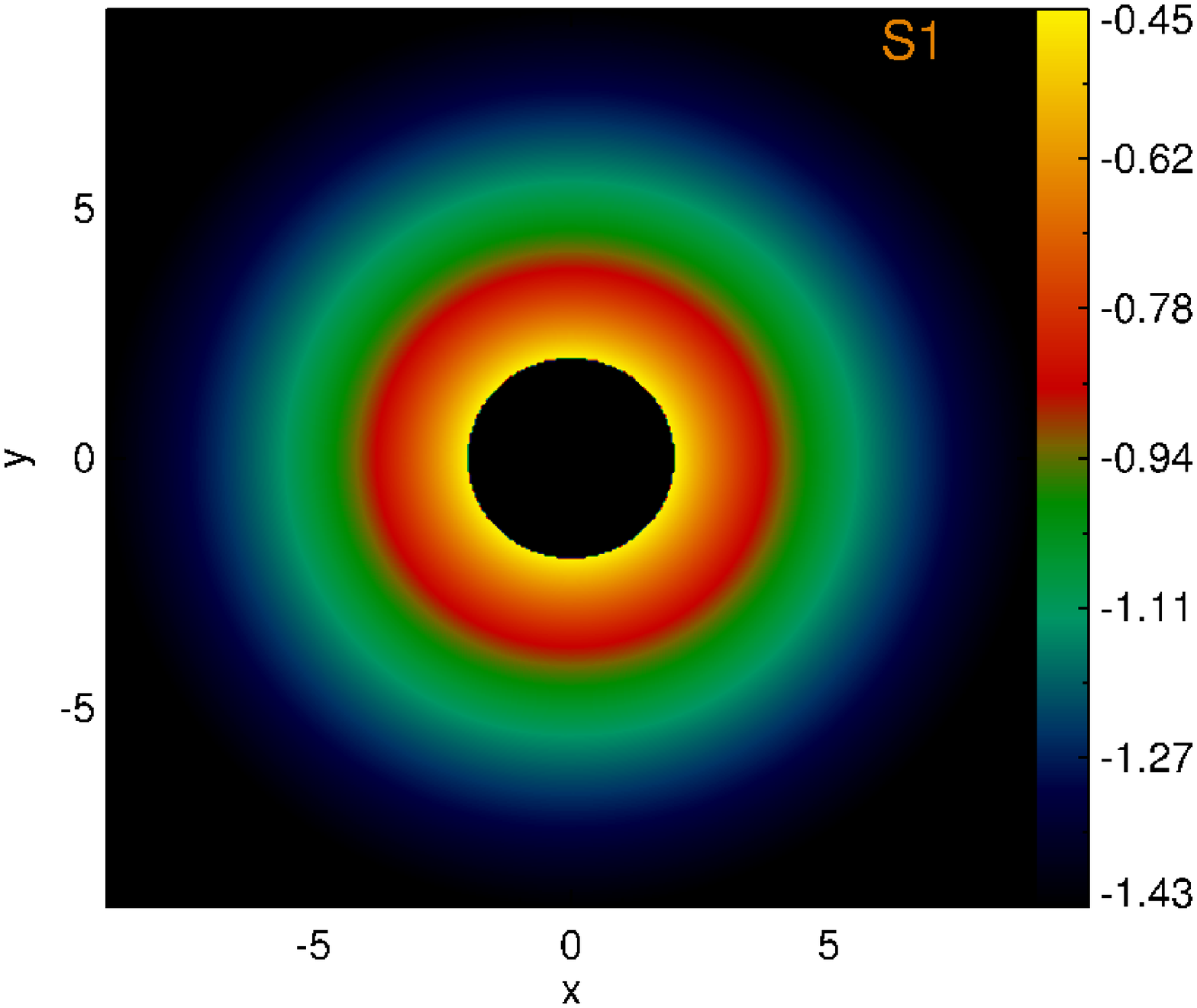}
  \includegraphics[width=8.5cm]{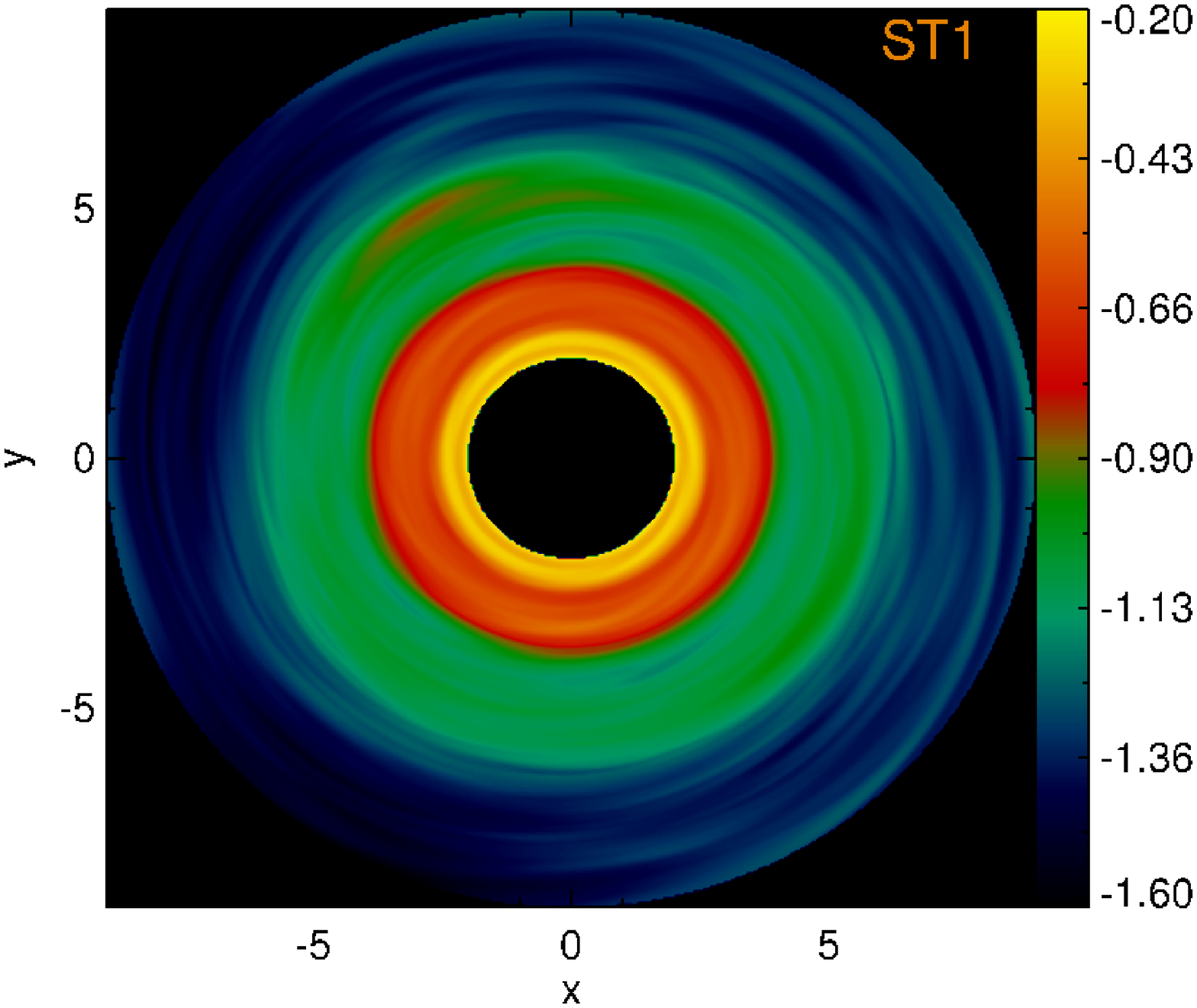}\\
  \includegraphics[width=8.5cm]{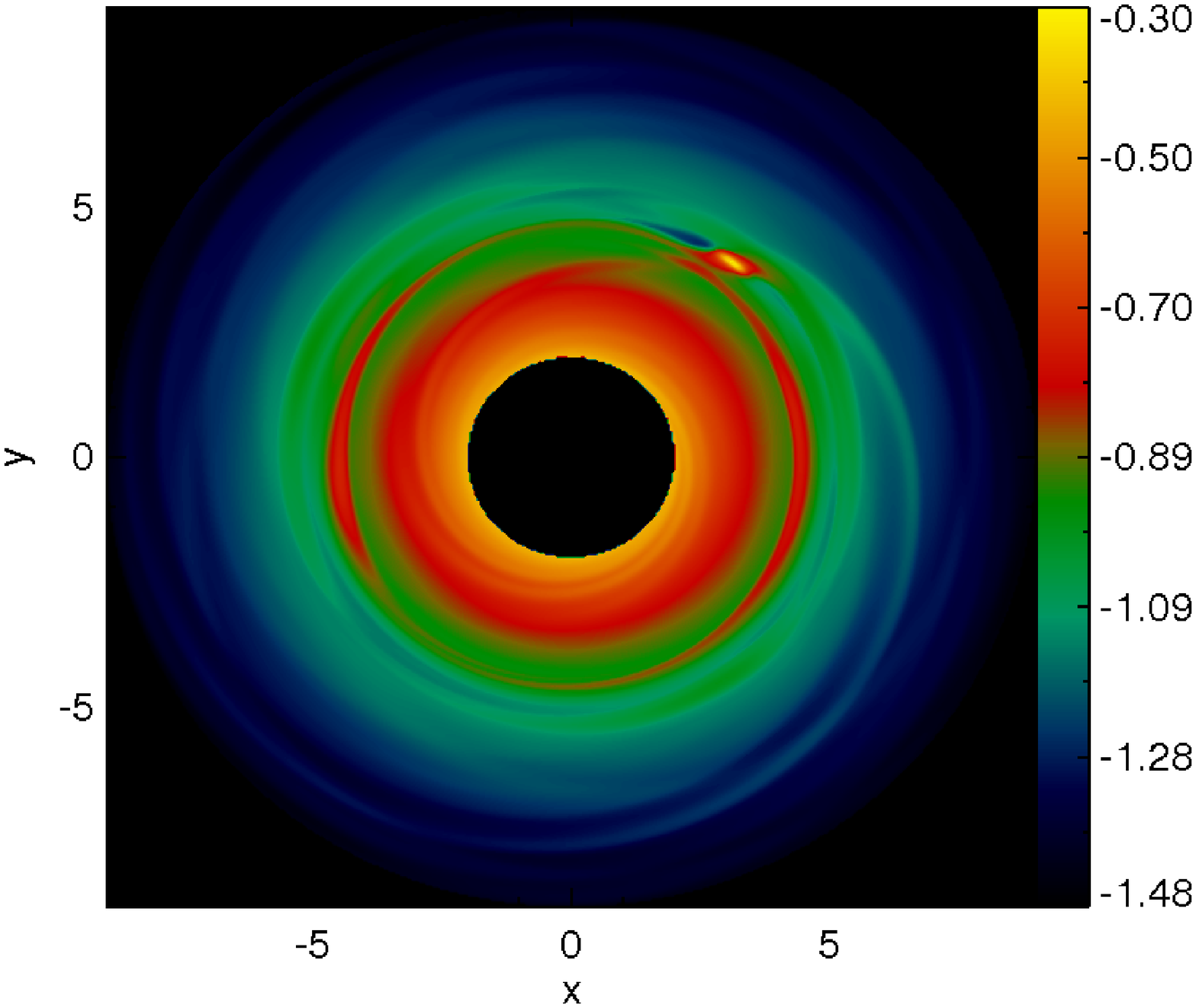}
  \includegraphics[width=8.5cm]{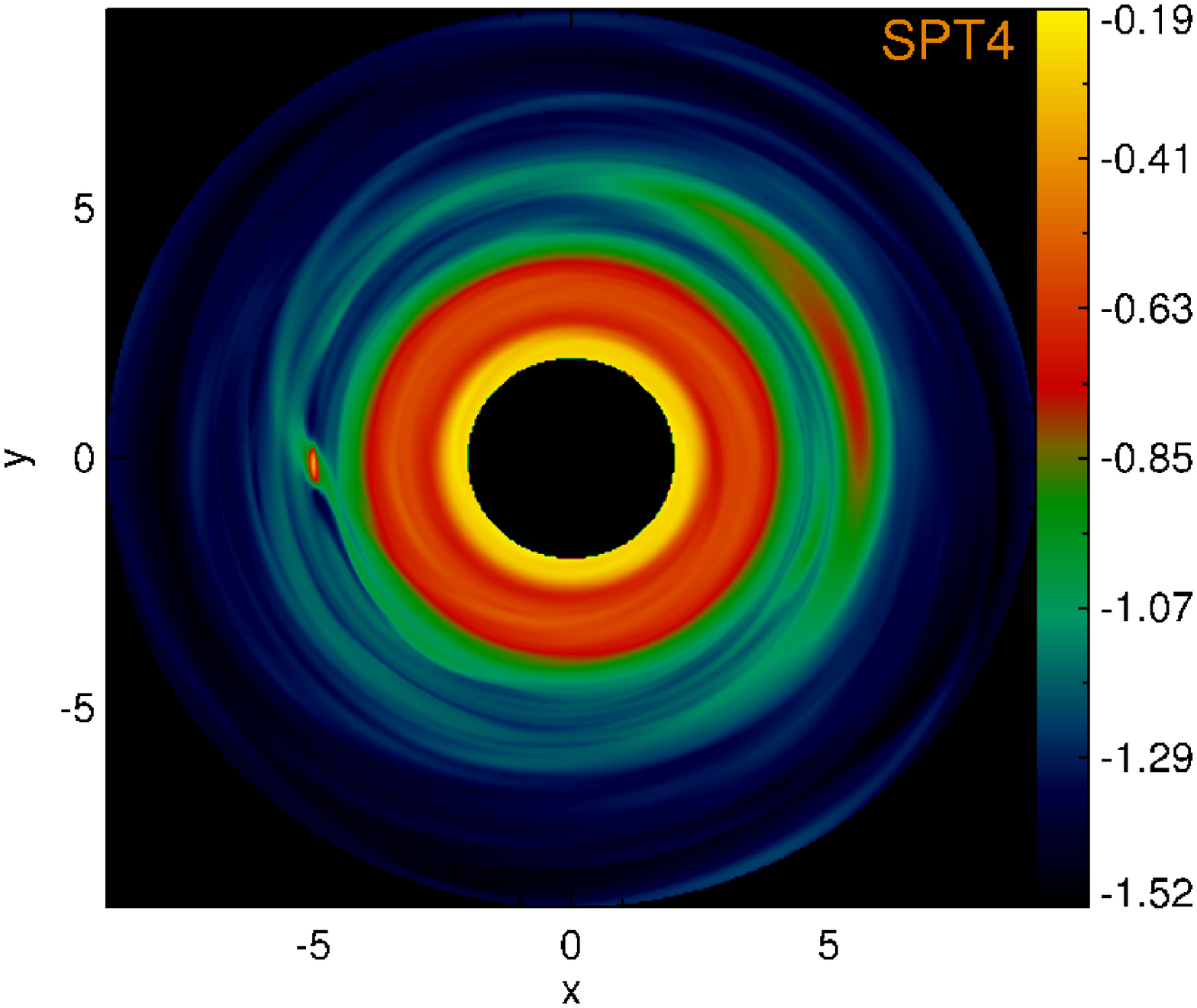}\\
  \includegraphics[width=8.5cm]{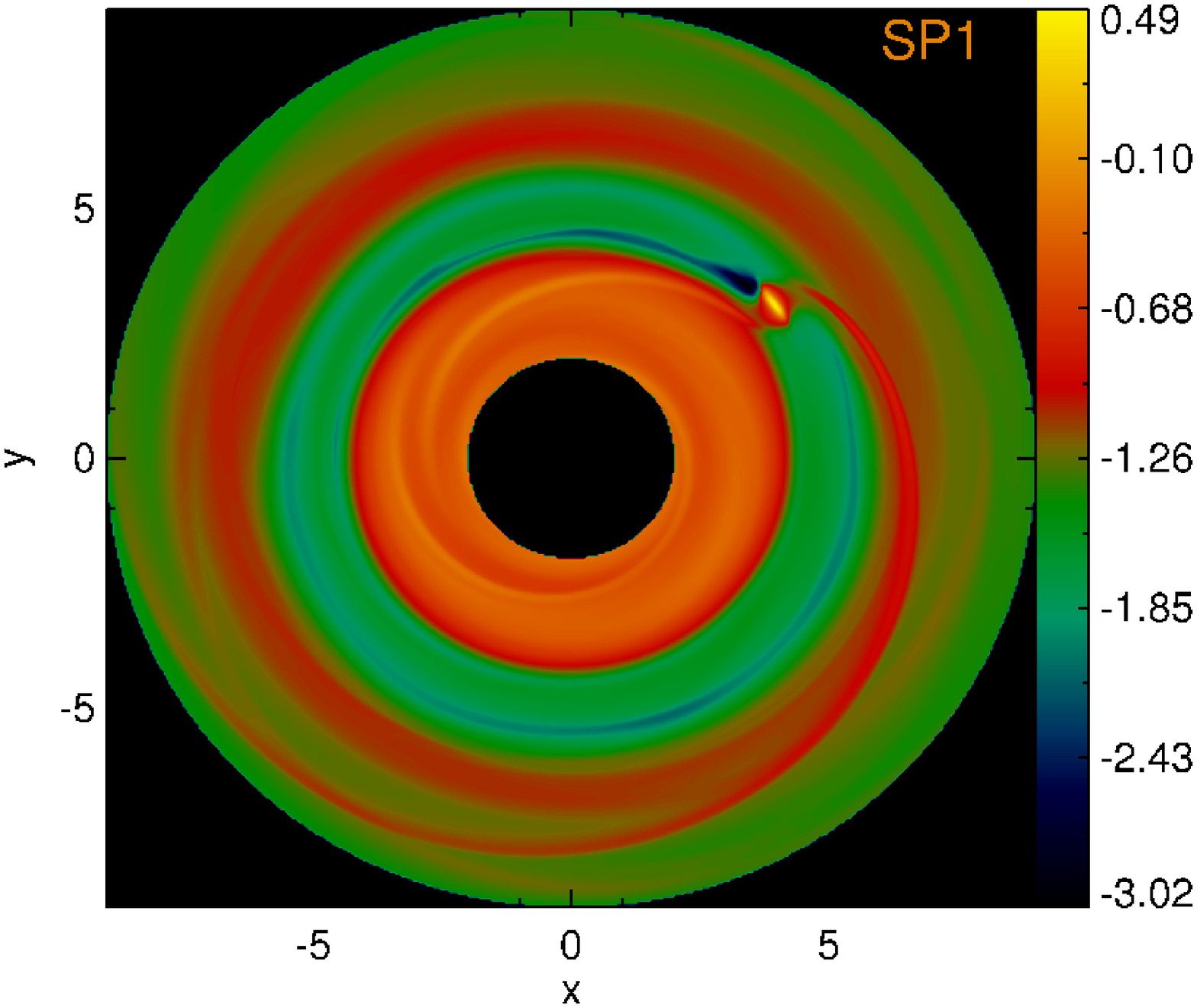}
  \includegraphics[width=8.5cm]{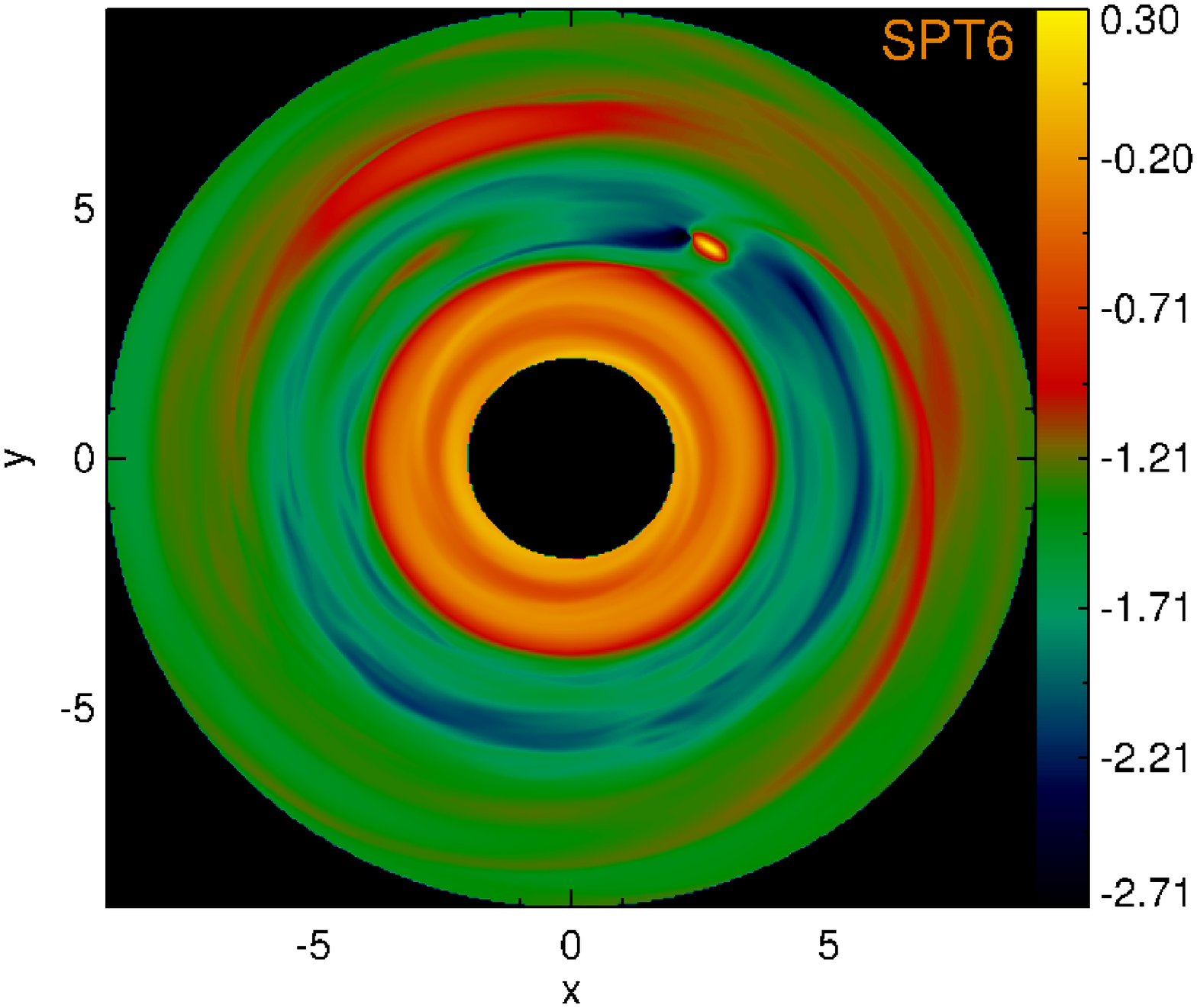}

\caption{Midplane density of all hydro- and magnetohydrodynamical disk models. The units of the axis are AU, and the midplane density is given in logarithmic scale of the code density unit. See Table \ref{tab:table_sims} for a detailed model description. \label{fig:f1}} 
\end{figure*}

\section{Simulation techniques}
 \label{sec:sim}
Motivated  by \citetads{2002ApJ...566L..97W} and \citetads{2005ApJ...619.1114W}, who investigated the observability of planet-induced structures in a selected sample of circumstellar disks, we now explore which properties of the central star, the disk, and the planet constrain clear detections. 
In the following, we briefly outline the applied hydrodynamical and the radiative transfer techniques. Thereafter, the observing simulations are diskussed briefly.\par

\subsection{Hydrodynamical simulation \label{sec:hydro}}

To simulate the interaction between a planet and a circumstellar disk, three-dimensional hydro- and magneto-hydrodynamical simulations were carried out using the finite volume fluid dynamics code PLUTO \citepads{2007ApJS..170..228M}. This computes Riemann fluxes using the HLLC and HLLD solvers for the hydrodynamic and magneto-hydrodynamic cases, respectively. Time integration is done using a second-order Runge Kutta scheme, and spatial interpolation is done using a second-order total variation diminishing (TVD) scheme. For the simulations that include magnetic fields, the constrained transport method for preserving a divergence-free magnetic field \citepads{2005JCoPh.205..509G} is used. Adiabatic index $\gamma = 1$ and isothermal approach is assumed, e.g., $T = T_0\, \left( r \sin{\vartheta}\right)^{-2b}$ see definition of speed of sound below.  
\par
The geometry of the computational grid is spherical; i.e., the coordinates are given as $(r,\vartheta,\varphi)$. The domain in code units are given by $r\in [1,10]$, $\vartheta \in [\pi/2 -0.3,\pi/2 +0.3]$ and $\varphi \in [0,2\pi]$. The grid resolution is $(N_\textup{r},N_{\vartheta},N_{\varphi})=(256,128,256)$, and it is centered on the center of mass of the planet-star system. 
\par
The gas disk is initially in sub-Keplerian rotation around the central star. The azimuthal velocity is given by
\begin{equation}
v_{\varphi}= \sqrt{v_\textup{k}^{2} - c_\textup{s}^{2}(a - 2b)},
\end{equation}
where $v_\textup{k}$ is the Keplerian velocity and $a=3/2$ and $b=0.5$ are the exponents of the radial power law distribution of the density $\rho\propto r^{-a}$ and sound speed $c_\textup{s}=c_{0}\, (r\sin\vartheta)^{-b}$. The initial density distribution is given by
\begin{equation}
\rho(r,\vartheta) = (r\sin\vartheta)^{-3/2}\exp\left(\frac{\sin\vartheta-1}{c_{0}^{2}}\right).
\label{eq:dens_prof2}
\end{equation}
The disk is described by a locally isothermal equation of state $P=c_\textup{s}^{2}\, \rho$. The ratio of the pressure scale height $h$ to the radial coordinate of the disk is taken to be a constant such that $h=H/(r\sin\vartheta)=0.07$. Therefore, the disk models are not flared. However, since flaring becomes important for disks predominantly heated by the central star, flaring will be considered in future studies. \par
The gravitational potential of the planet is given by a softened point-mass potential
\begin{equation}
\Phi_\textup{p}(\mathbf{r}) = -\frac{GM}{(|\mathbf{r}-\mathbf{r}_{p}|^2 + \epsilon^{2})^{1/2}}
\end{equation}
where $\epsilon$ is the softening parameter, needed to avoid very high values of the potential near the planet position that could induce very high velocities and numerical problems. For all the simulations $\epsilon$ is set to be a fraction of the Hill radius $\epsilon=l \cdot r_\textup{p}(M_{p}/3)^{1/3}$ with $l=0.3$. 
\par
The vertical extent of the simulations is $\vartheta \in [\pi/2 -0.3,\pi/2 +0.3]$, since for the purpose of the hydrodynamical evolution, there is no need to simulate the entire spherical domain $\vartheta \in [0,\pi]$, because this would only complicate the simulations and significantly increase the computational time. The simulated domain covers four pressure scale heights above and below the midplane of the disk. For the purpose of the radiative transfer analysis, however, it is necessary to have the full $\vartheta$ domain, so as to follow the radiation at all possible angles. For this reason, the domain is increased a-posteriori, providing the full $\pi$ range in the $\vartheta$ direction (see \S \, \ref{sec:radi}). In the extended domain, the density in the disk is set to zero, which is a good approximation, since the density at four pressure scale heights has decreased several orders of magnitude over the midplane density. 
\par
Additionally, since in the magnetohydrodynamical simulations, the zones $r<2$ and $9<r<10$ are boundary buffer zones where resistivity is applied, these parts of the domain are generally \textit{not} included in the radiation transport analysis. This means that the effective grid for the radiative transfer is $r\in [2,9]$ (in PLUTO code units). The number of cells in PLUTO is $(N_\textup{r},N_{\vartheta},N_{\varphi})=(200,128,256)$.
\par
Different configurations were explored. The planet mass $M_\textup{p}$ to stellar mass $M_\star$ ratio is denoted as
\begin{equation}
\eta=\frac{M_\textup{p}}{M_\star}.
\end{equation}
A list of the simulations performed and of the parameters used in each of them is given in Table \ref{tab:table_sims}.
Our setup contains six simulations in all. First we performed a simulation of an unperturbed disk to compare the other simulations with. Then we focused on planets with $\eta=0.001$ to investigate the observability of Jovian-mass objects in circumstellar disks. Furthermore, we selected planets with $\eta=0.0001$ to estimate the possibility of tracing lower mass objects with ALMA; i.e., the mass of a planet depends on the mass of the star.  Additionally, all simulations were carried out for the hydro- and the magnetohydrodynamical case to determine the influence of magnetic fields on the observability of large-scale structures. 
Local and global magnetohydrodynamical simulations of accretion disks without the presence of a planet have shown the formation of zonal flow features, which are axis symmetric density enhancements in the disk confirmed in modifications of the rotation law (see Fig. \ref{fig:f1} top, left). These features are effects of the non locality of the Maxwell stresses associated with MRI. These zonal flows are assumed to be of major importance for the accumulation of dust: \textit{1.)} retain particles for a long time in the disk \citepads{2012A&A...545A..81P}, \textit{2.)} to create large enough gas-to-dust ratios to trigger planetesimals formation \citepads{2009ApJ...697.1269J}.\par 
 
\subsection{Radiative transfer \label{sec:radi}}
\paragraph{Continuum radiative transfer:} The follow-up radiative transfer simulations were performed with the Monte Carlo 3D continuum radiative transfer code MC3D (\citealtads{1999A&A...349..839W}; \citealtads{2003CoPhC.150...99W}), which self-consistently calculates the dust temperature of an arbitrary density distribution. Based on this, the resulting spectral density distributions (SED), re-emission, and scattered light maps were simulated. Besides analytically defined dust density profiles, MC3D allows reprocessing a numerically generated density structure, such as those resulting from the hydrodynamical simulations. However, the outcome of the simulations with PLUTO has to be converted for the use in MC3D.

\paragraph{Conversion and input of hydrodynamical data in MC3D:} By merging eight neighboring cells from the PLUTO grid into one we significantly decrease the required computing time, which is essential given the large parameter space (see \S \, \ref{sec:model}). The corresponding diskrepancies between a high resolution re-emission image and a merged one are lower than 0.1\%.
Additionally, the PLUTO code computes its results in a cut-out in the $\vartheta$-direction around the disk midplane (see \S \, \ref{sec:hydro}). This is reproduced by assuming a density equal to zero in the part of the model space that was not simulated with PLUTO (\S \, \ref{sec:hydro}). Furthermore we use as many cells in $\vartheta$-direction as needed to get the same resolution in the cut-out area. The grid structure of MC3D is $(N_\textup{r}, N_\vartheta, N_\varphi) = \left(100, 341, 128\right)$ resulting in a total cell number of $4.36 \cdot 10^6$. The images simulated in the radiative transfer are fixed to a resolution of $201 \, \textup{px} \times 201 \, \rm px$. This ensures that the synthesized beam of ALMA in the simulated observations always contains several tens of pixels (see \S \, \ref{sec:noise}).

\begin{figure}[tp]
  \resizebox{\hsize}{!}{\includegraphics{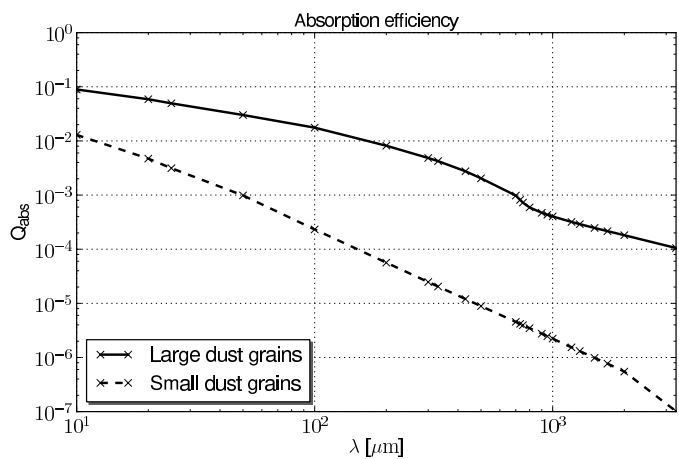}}
\caption{Absorption efficiency for the large and small dust grains in the mid-infrared to mm-regime. The solid line represents large dust grains and the dashed line small dust grains. For further information see \S \, \ref{sec:radi}. All plots were created with \textit{python matplotlib} \citepads{Hunter:2007}. \label{fig:qabs}}
\end{figure}

\begin{figure}[t]
  \resizebox{\hsize}{!}{\includegraphics{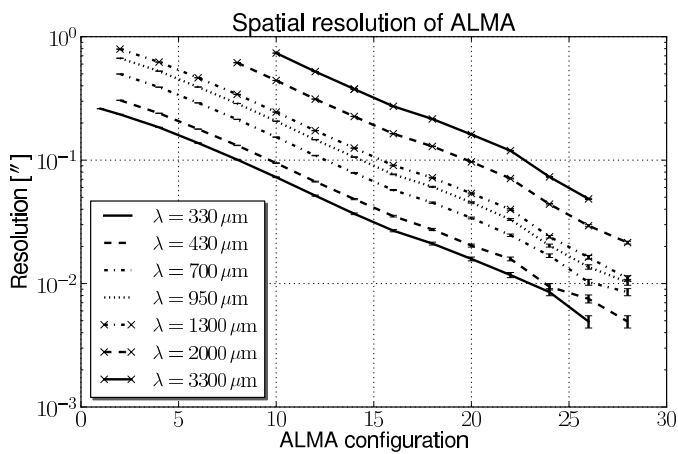}}
\caption{Resolution of ALMA at the seven wavelengths considered. Different line styles represent different wavelengths. These values are derived from the synthesized beams simulated with CASA. \label{fig:resolution}}
\end{figure}

\paragraph{Dust properties:} It is assumed that the gas and the dust within the circumstellar disk are perfectly mixed with a gas-to-dust mass ratio of $100:1$.
The dust grains are spherical and consist of $62.5\, \%$ silicate and $37.5\, \%$ graphite (optical data by \citealtads{2001ApJ...548..296W} and concept by \citealtads{1977ApJ...217..425M}). Ice were not considered in the current the study, but it will be regarded on our future investigations because its impact in particular in the outer parts of circumstellar disks is assumed not to be negligible. The grain size distribution follows a power law, $n(a) \propto a^{-q}$ with an exponent $q = 2.5$ \citepads{1969JGR....74.2531D}. We selected two ranges for the dust grain radius $a$.
The range ($a \in \left[0.005\, \mu \textup{m}, 100 \, \mu \textup{m}\right]$) mimics the larger grain size frequently found in the denser regions around the midplane of a circumstellar disk (\citealtads{2003ApJ...588..373W}; \citealtads{2012A&A...543A..81M}). Since giant planets induce structures, in particular in this region of a disk, it is necessary to consider this dust species in our study. The other dust grain radius range ($a \in \left[0.005\, \mu \textup{m}, 0.25 \, \mu \textup{m}\right]$) represents the interstellar medium \citepads{1979ARA&A..17...73S}, which is important for this study because some young stellar objects \citepads[e.g. CB26,][]{2009A&A...505.1167S} show a lack of large dust particles.\par
The absorption properties of the dust depend on the grain size distribution.
For comparison, the absorption efficiency $Q_\textup{abs}$ calculated using Mie theory for both grain sizes is shown in Fig. \ref{fig:qabs}. 
The grain size distribution can be approximated by one effective dust grain radius $a_\textup{eff}$ resulting from the volume averaging:

\begin{equation}
 a_\textup{eff} = \left( \left(\frac{1 - q}{4-q} \right) \, \left(\frac{a_\textup{max}^{4-q} - a_\textup{min}^{4-q}}{a_\textup{max}^{1-q} - a_\textup{min}^{1-q}} \right) \right)^{\frac{1}{3}}. \label{glg:aeff}
\end{equation}
Based on this, the mass density from the hydro- and magnetohydrodynamical simulations is transformed into a number density as needed by the MC3D.
The planetary heating of the circumplanetary environment, considered already by \citetads{2005ApJ...619.1114W}, and heating processes by viscosity and accretion onto the star or the planet are not taken into account in this study. The only heating source is the photosphere of the central star in a blackbody approximation, therefore the disks are heated passively.

\subsection{(Sub)mm-observations: Simulation setup \label{sec:noise}}

\begin{figure}[t]
  \resizebox{\hsize}{!}{\includegraphics{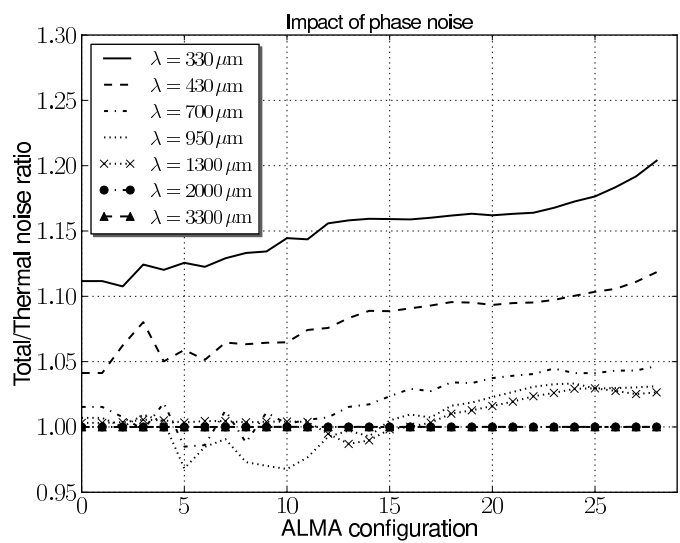}}
\caption{Total-to-thermal noise ratio. For the phase noise calculation we use a PWV$ = 1.0 \, \rm mm$ with a deviation of $\Delta$PWV $= 0.15 \, \rm mm$, with an exposure time of two hours. Different line styles represent different wavelengths. Thanks to the sequential simulation process of thermal and phase noise, ratios below 1.00 are possible, because the phase and thermal noise can erase each other. \label{fig:inf_noise}}
\end{figure}

\paragraph{ALMA:} After its completion ALMA will consist of 66 (sub)mm-antennas (54 $\times$ 12-m- and 12 $\times$ 7-m-antennas), which can be arranged in 28 different array configurations. ALMA will be able to observe in ten bands between $0.3\, \rm mm$ and $9.3\, \rm mm$ \citepads{2004AdSpR..34..555B}.

\paragraph{CASA:} We simulated observations of a circumstellar disk for every second ALMA-antenna configuration, by using the CASA 3.2 simulator \citepads{2012ASPC..461..849P}. The number of considered bands/wavelengths amounts to seven (shown in Tab. \ref{tbl:wavel}).
The angular resolution of ALMA is shown in Fig. \ref{fig:resolution} as a function of wavelength and array configuration. See appendix \ref{a:alma} for further information on the interferometer.  
Furthermore, we consider three different exposure times ($\frac{1}{2}$h, $2$h, $8$h).

\begin{table}[htbp]
\caption{Selected wavelengths for the simulated ALMA observations. \label{tbl:wavel}}
\begin{center}
\begin{tabular}{ccc}
\hline\hline
Central wavelength $\lambda$  & Central frequency  $\nu$ & ALMA band\\
 {[$ \rm \mu m$]} & {[$\rm GHz$]}  &       \\
\hline
330 & 910 & 10 \\
430 & 700 & 9 \\
700 & 430 & 8 \\
950 & 320 & 7 \\
1300 & 270 & 6 \\
2000 & 150 & 4 \\
3300 & 90 & 3 \\
\hline
\end{tabular}
 \tablefoot{Bandwidth: $8 \, \rm GHz$}
\end{center}
\end{table}
The simulated observations include the significant influence of thermal noise, which is caused by precipitable water vapor (PWV) in the Earth's atmosphere. For the wavelength-dependence of the PWV we follow the CASA cookbook \citepads{casa} and consider the recommended values.\par
From simulated observations of an empty patch of the sky, \textit{noise maps} are derived to obtain the impact of thermal noise on our setup. In addition, the CASA simulator is also able to simulate phase noise.
Because of the high computing time requirement of phase noise simulations, we first explored the impact of phase noise on the image quality by adding it to the already calculated \textit{thermal noise maps} (we call the outcome total noise). The total-to-thermal noise ratio is plotted in Fig. \ref{fig:inf_noise}. It only shows significant deviations ($> 10 \%$ for every ALMA configuration) for the shortest wavelength.\par 
In another test, 10\% of our parameter space was simulated again, considering thermal and phase noise.
In comparison to the case with thermal noise only, just 0.6\% of the simulated observations are evaluated differently (for assessment criteria see \S \, \ref{sec:analysis}). For this reason we decide to consider only thermal noise for our study.\par
\textit{Noise maps} and all simulated observations were calculated for the Taurus-Auriga star forming region considering the position of TTAU ($\alpha =$ 04h22m59s, $\delta = $+19$^\circ$32$\arcmin$06$\arcsec$, J2000), which corresponds to a maximum elongation of $\approx 50^\circ$ in December.
On the smallest scales of the spatial resolution of ALMA it can be necessary to figure the observation out with mosaic imaging if the maximum angular scale of ALMA is smaller than the object.

\section{Model setup}
\label{sec:model}

\begin{figure}[htp]
  \resizebox{\hsize}{!}{\includegraphics{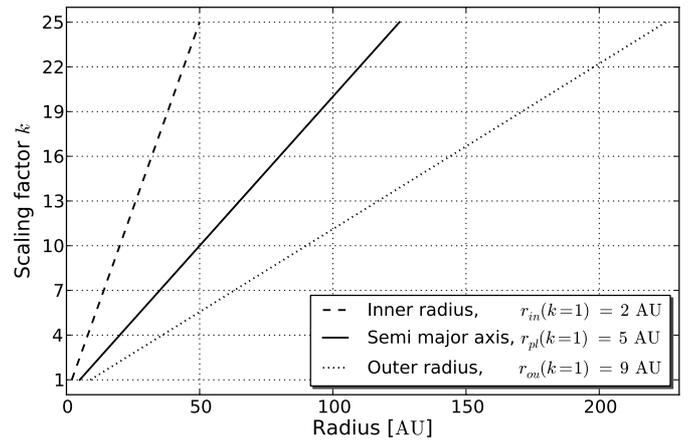}}
\caption{Correlation between scaling factor $k$ and the inner (dashed line) and outer disk (dotted line) radius ($r_\textup{in}, r_\textup{out}$) and the semi-major axis (solid line) of the planet ($r_\textup{pl}$). \label{fig:k-R}}
\end{figure}

\paragraph{Disk size:} The self-similar hydro- and magnetohydrodynamical simulations allow us to arbitrarily scale the size and the total mass of the disk.
The disk size is adapted by scaling the inner and outer disk radius linearly with a factor $k \cdot \rm AU$ in the range of $k \in \left\{1,4,7,10,13,16,19,22,25\right\}$.
In the case of $k = 1$, the hydrodynamical simulations are used unscaled. The maximum value $k = 25$ was chosen to simulate a planet with a semi-major axis ($125 \, \rm AU$) similar to Fomalhaut b \citepads{2008Sci...322.1345K}. For illustration the inner and outer disk radius and the semi-major axis of the planet are shown as a function of $k$ in Fig. \ref{fig:k-R}. In a more general view, our disk models are anulli or smaller cut-outs from large circumstellar disks. However, since we investigate the ability to trace structures of a defined size in our disk models, it is not necessary to take more disk areas into account. In the following, we call these models \textit{disks}.
\paragraph{Total disk masses:} The typical estimates of the total mass of circumstellar disks is in the range of $0.01 \, \rm M_\odot$ to $0.1 \, \rm M_\odot$ \citepads{2010ApJ...723.1241A}.
During the disk evolution, the disk mass decreases because of, i.e., photoevaporation and accretion \citepads{2012A&ARv..20...52W}. Half of all circumstellar disks have a lifetime of $\leq 3\, \rm Myr$ \citepads{2001ApJ...553L.153H}. To consider this factor in our current study, a wide range of masses was taken into account. 
At the lowest disk masses in our setup, which mimic the mass typical of a debris disk, our hydro- and magnetohydrodynamical disk models may no longer correcty approximate the disk structure, because additional processes, e.g. radiation pressure and photoevaporation, become important. Nevertheless, observations were simulated to predict the minimum mass that allows one a detection of a circumstellar disk.
The following disk masses are considered (only integers in the exponent):
\begin{equation}
 M = 2.67 \cdot 10^{-2\, \dots\, -7} \, \rm M_\odot.
\end{equation}
The value $2.67$ results from the volume integration of the density structure of the disk models from the hydro- and magnetohydrodynamical simulations. As an alternative to this approach, we prepared a setup for a total disk mass derived from the disk of the Butterfly Star (\citealtads{2003ApJ...588..373W}) in \S \, \ref{sec:butter}.

\paragraph{Distance, stars, and disk inclination: \label{sec:stars}}
Throughout the entire study objects are located at a distance of $140 \, \rm pc$ (typical for TT-objects). We assume disks that are seen almost face-on (inclination angle $i = 5^\circ$).
The number of different central stars, characterized by luminosity and effective temperature, amounts to six. Table \ref{tbl:stars} summarizes the important stellar parameters for the four main-sequence stars and two typical pre-main-sequence stars.
\begin{table}[h]
\caption{List of the selected stars. \label{tbl:stars}}
\centering
\begin{tabular}{lccc}
\hline \hline
Spectral type & L & T$_\textup{eff}$ & M \\
  & {[L$_\odot$]}  &	{ [K]}	& {[$\textup{M}_\odot$]} \\
\hline
K (MS)& 0.35 & 4500 & 0.7\\
G (MS)& 1 & 6000 & 1.0\\
F (MS)& 7.5 & 6900 & 1.8\\
A (MS)& 20 & 8500 & 2.4\\
 \hline 
\multicolumn{1}{l}{T-Tauri (PMS)} & 0.95 & 4000 & 0.5 \\
\multicolumn{1}{l}{Herbig Ae (PMS)} & 43 & 9500 & 2.5\\
\hline
\end{tabular} 
 \tablefoot{The denotation and the effective temperature of the main-sequence stars as spectral types refers to the Harvard Classification of Stars. We use the mass-luminosity relationship to estimate the stellar mass. (MS) = main-sequence stars; (PMS) = pre-main-sequence stars.}
\end{table}

\section{Analysis and data reduction}
\label{sec:analysis}

\begin{figure}[t]
  \resizebox{\hsize}{!}{\includegraphics{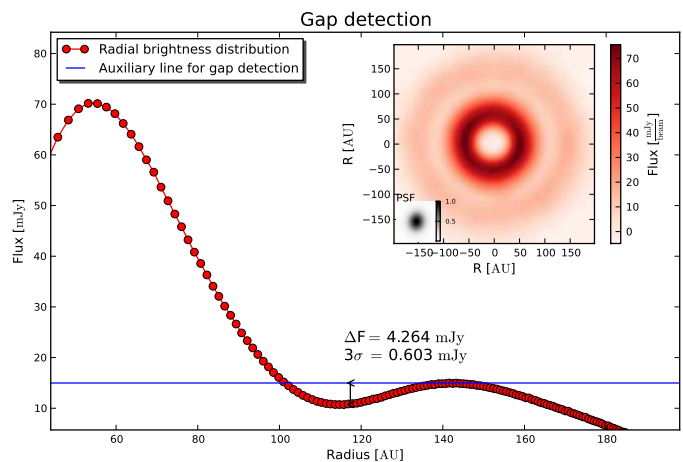}}
\caption{Illustration of the radial brightness distribution (solid line) with auxiliary line (dotted line) used for the automatical gap detection. The corresponding simulated observation is shown in the upper right. The disk model has a scaling factor of $k = 22$ (outer disk radius $198\, \rm AU$), a total disk mass of $M = 2.67 \cdot 10^{-2} \, \rm M_\odot$, and a \textit{G type star} (\S \, \ref{sec:stars}). We used the 8$^{\rm th}$ ALMA configuration and a wavelength of $430 \, \rm \mu m$. A gap will be detected if the difference between the brightness profile and the auxiliary line at the local minimum of the brightness profile is three times larger than the standard deviation of the noise of the observation (noise: see \S \, \ref{sec:noise} ). \label{fig:pladet}}
\end{figure}
We now briefly diskuss the computational analysis of the output from the CASA simulator. There are three steps for the assessment of every simulated observation. First we derive the standard deviation of a simulated observation from the corresponding \textit{noise maps} (\S \, \ref{sec:noise}). If the signal-to-noise ratio is greater than three, this simulated observation will be taken into account for further analysis.\par
In the next step, the object is considered as \textit{resolved} if the area of the synthesized beam is smaller than the object in the simulated observation image in a $3\sigma$ environment. This analysis is sufficient for the unperturbed disk model, but the other disk models have to fulfill a criterion that indicates the presence of large-scale disk structures.\par
The most prominent structure a giant planet induces in the disk is a gap \citepads{1984ApJ...285..818P}. Because of the radial symmetry of the gap, it is unequivocally detectable in the radial brightness profile.
In our setup, the disk's density models show a significant amount of material in the inner part of the disk (within the orbit of the planet). The radial brightness profile indicates this by a global maximum of the brightness. In the low-density area of the gap, the brightness profile decreases towards a local minimum, followed by another local maximum. We fit a line, parallel to the abscissa, onto this local maximum (see Fig. \ref{fig:pladet} for illustration). Now, a gap will be assessed as \textit{detected}, if the difference between this auxiliary line and the local minimum is larger than three times the standard deviation derived from the \textit{noise maps} ($3\sigma$ environment). This order is chosen deliberately, because a simulated observation that does not meet a previous condition also does not meet all the following ones.\par
\begin{figure*}[!ht]
\begin{center}
\normalsize{\textbf{$\boldsymbol{\lambda = 430\, \rm \mu m}$, large dust grains, unperturbed disk (S1)}}\\ \end{center}
\centering
  \includegraphics*[width=17cm]{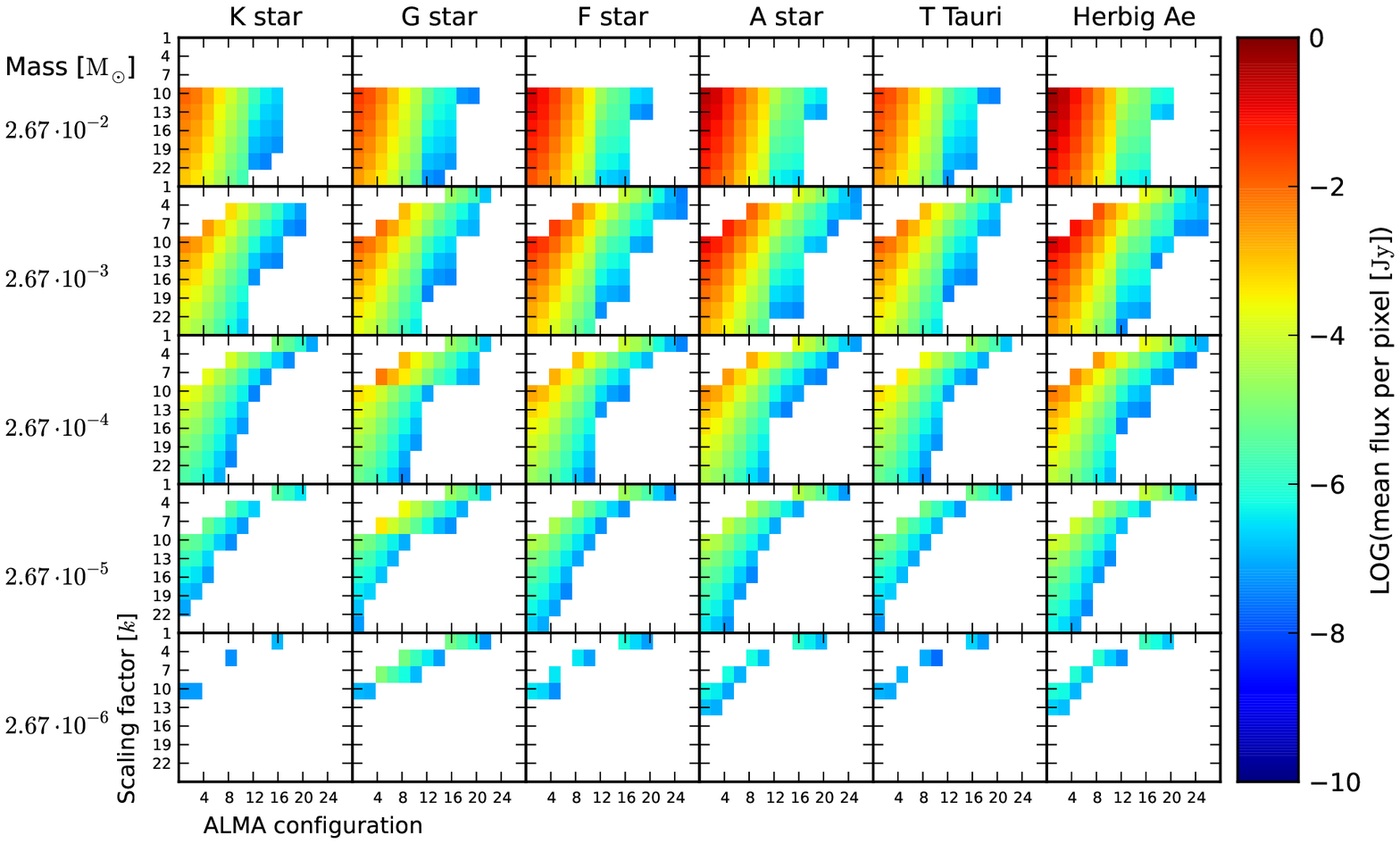}\\
\begin{center}
\normalsize{\textbf{$\boldsymbol{\lambda = 430 \, \rm \mu m}$, large dust grains, planet with $\boldsymbol{\eta=0.001}$ in the disk (SP1)}}\\ \end{center}
\centering
  \includegraphics*[width=17cm]{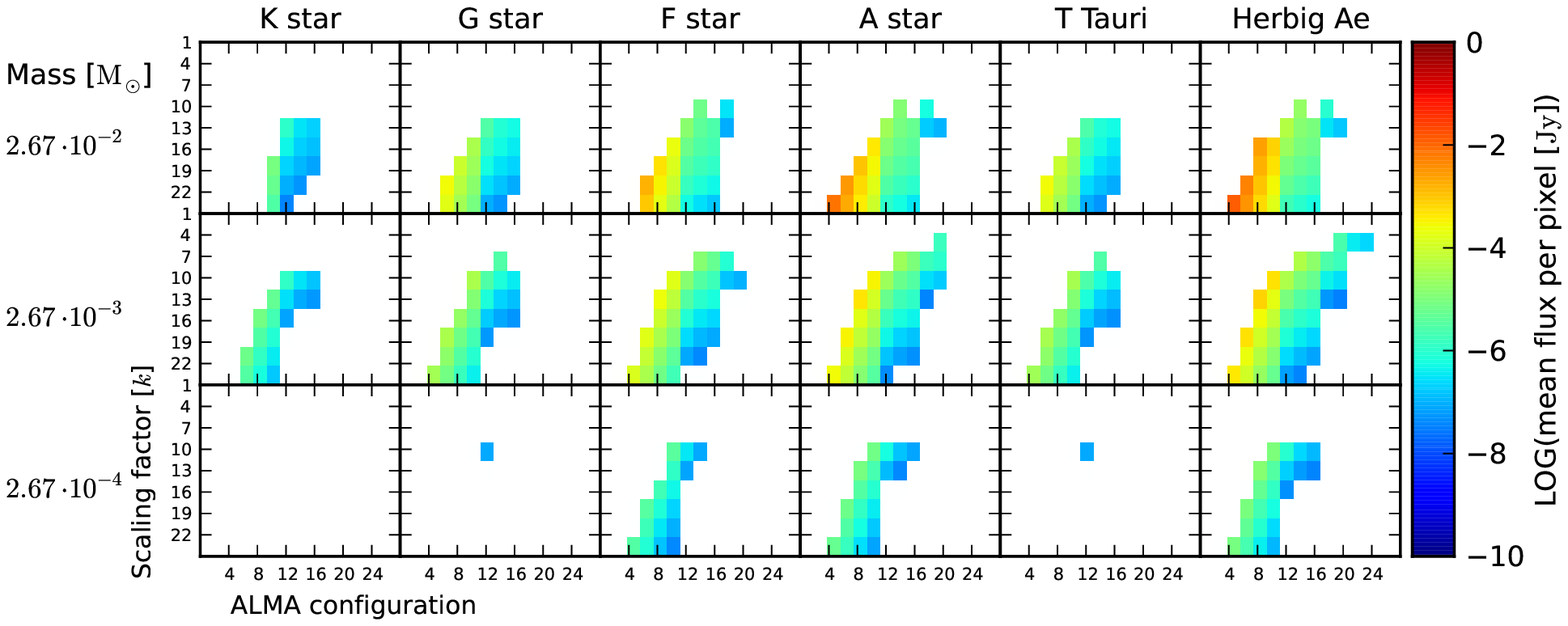}\\
\caption{Feasibility of spatially resolving an unperturbed disk (top) and of detecting a gap induced by a planet with a planet-to-star mass ratio of $\eta=0.001$ (bottom) depending on the disk mass, the host star, the disk size, and ALMA configuration. The disks are made up of large dust grains. The exposure time is two hours at a wavelength of $430\, \mu \rm m$. The simulations for a combination of a total disk mass of $M = 2.67 \cdot 10^{-2} \, \rm M_\odot$ and a scaling factor $k < 10$ were not calculated. For the scaling factor $k$ see Fig. \ref{fig:k-R}, for the resulting angular resolution depending on wavelength and ALMA configuration see Fig. \ref{fig:resolution}. The scale of one pixel is $0.00064\arcsec \times k$. \label{fig:sp1r28all_phase62a}}
\end{figure*}
\begin{figure*}[!htp]
\begin{center}
\normalsize{\textbf{$\boldsymbol{\lambda = 430 \, \rm \mu m}$, small dust grains, planet with $\boldsymbol{\eta=0.001}$ in the disk (SP1)}}\\ \end{center}
\vspace{-0.4cm}
\centering
  \includegraphics*[width=17cm]{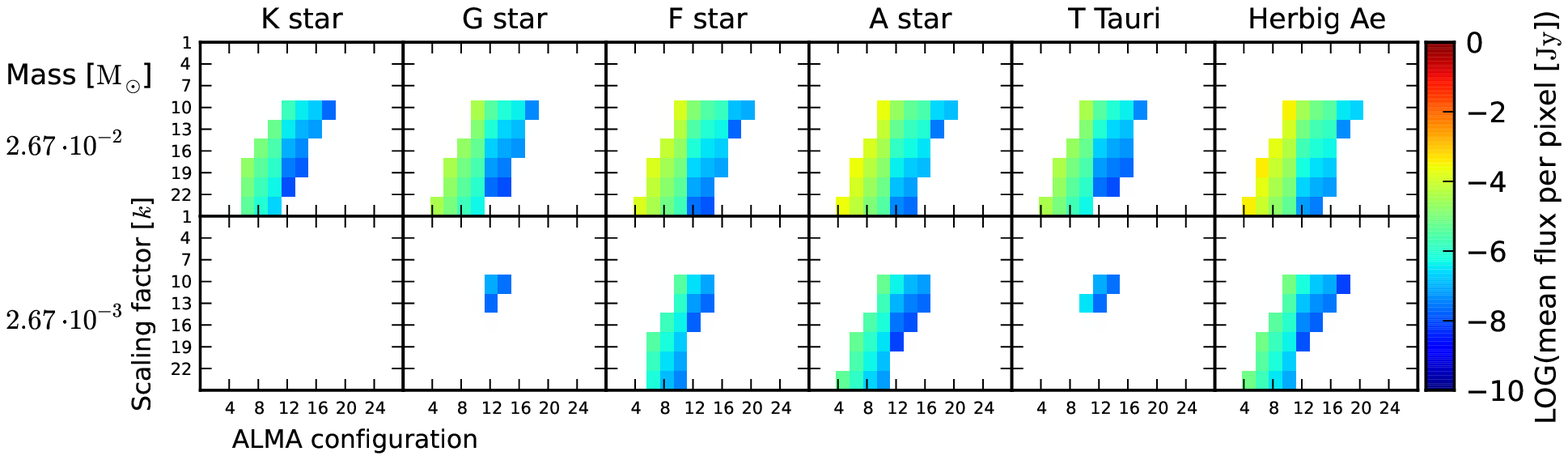}\\
\vspace{0.4cm}
\begin{center}
\normalsize{\textbf{$\boldsymbol{\lambda = 1300 \, \rm \mu m}$, large dust grains, planet with $\boldsymbol{\eta=0.001}$ in the disk (SP1)}}\\ \end{center}
\centering
\vspace{-0.2cm}
  \includegraphics*[width=17cm]{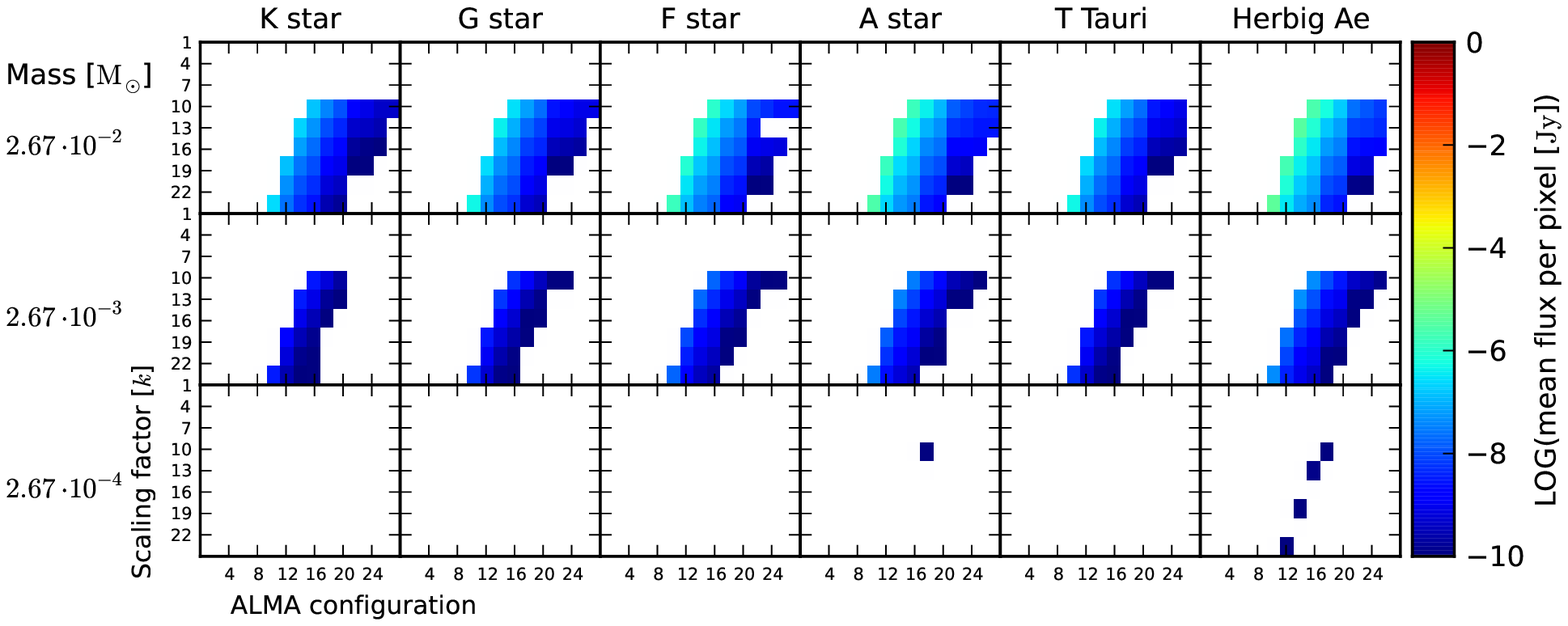}\\
\vspace{0.4cm}
\begin{center}
\normalsize{\textbf{$\boldsymbol{\lambda = 330 \, \rm \mu m}$, large dust grains, planet with $\boldsymbol{\eta=0.001}$ in the disk}}\\ \end{center}
\centering
\vspace{-0.4cm}
  \includegraphics*[width=17cm]{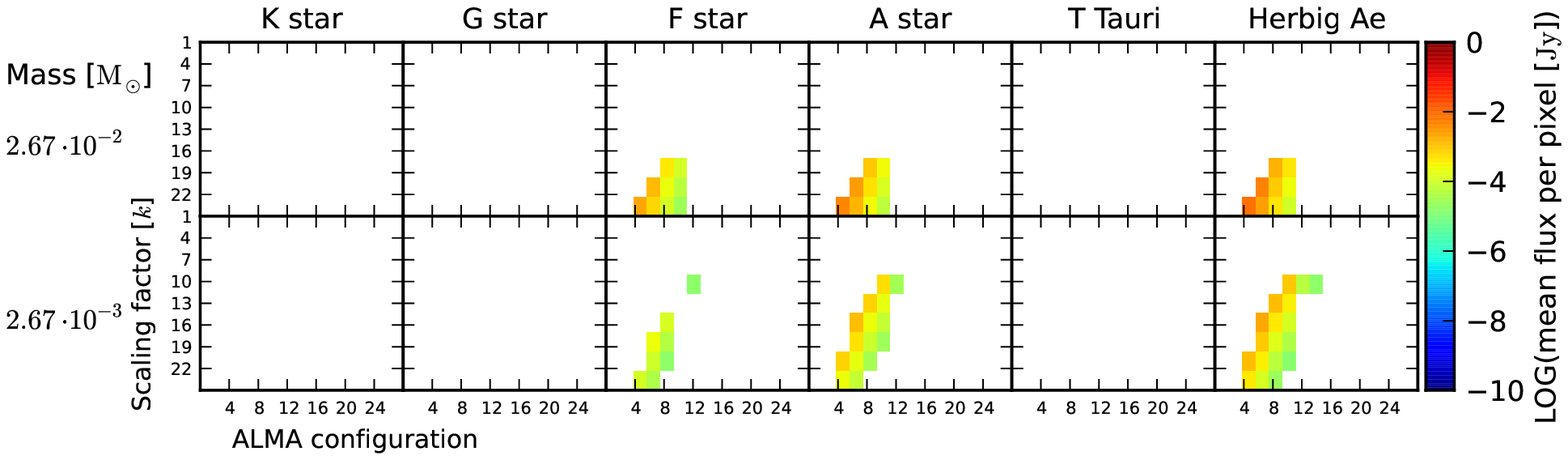}

\caption{Wavelength dependence of the detectability of a gap induced by a planet with $\eta=0.001$ in a circumstellar disk depending on the disk mass, the host star, the disk size, and the ALMA configuration. Both grain species are considered. The assumed exposure time amounts to two hours. The simulations for a combination of a total disk mass of $M = 2.67 \cdot 10^{-2}\, \rm M_\odot$ and a scaling factor $k < 10$ were not calculated. For the scaling factor $k$ see Fig. \ref{fig:k-R}; for the resulting angular resolution depending on wavelength and ALMA configuration see Fig. \ref{fig:resolution}. The scale of one pixel is $0.00064\arcsec \times k$. \label{fig:sp1r28all_phase62b}}
\end{figure*}

The comprehensive illustration of the statistical outcome of the study is given in overview charts (see Fig. \ref{fig:sp1r28all_phase62a}, \ref{fig:sp1r28all_phase62b}). They summarize the impact of the stellar type, the total disk mass, the disk size, and ALMA configuration on the observability in just one figure. If a simulated observation fulfills the criteria mentioned above, the mean flux per pixel of the simulated image will be drawn graycolor-coded (in the online-version color-coded) into the figure. A colorbar indicates the value of the flux per pixel on a logarithmic scale. The presentation is only made for disks with masses that allow for an observation in at least one setup.

\section{Results}
\label{sec:results}

We now present the main results from our studies of the observability of planet-induced structures in circumstellar disks. First, the feasibility of spatially resolving the various disks and cut-outs from disks is diskussed. Thereafter we explore the detectability of planet-induced gaps under the influence of the individual parameters on our setup. Based on this, we predict the optimal parameter combination to observe a selected object with ALMA. Additionally, we propose a survey of nearby, circumstellar disks seen face-on. For a general overview, Figs. \ref{fig:sp1r28all_phase62a} to \ref{fig:sp1r28all_phase62b}  show a selection of five overview charts (see \S \, \ref{sec:analysis}).
In comparison, the results for an unperturbed disk are shown in Fig. \ref{fig:sp1r28all_phase62a} at the top. Thereafter, the outcome for a disk with a planet with $\eta=0.001$ is drawn. Until \S \, \ref{sec:mhd} only the hydrodynamical case is explored.

\subsection{Feasibility of spatially resolving circumstellar disks}
Our investigations based on the unperturbed disk model (S1, see \S \, \ref{sec:hydro}) outline that circumstellar disks at a distance of $140\, \rm pc$ can be spatial resolved with ALMA around all stars, disk sizes, and at all wavelengths in our setup. Even observations of disks with masses down to $2.67 \cdot 10^{-6}\, \rm M_\odot$ are feasible (see Fig. \ref{fig:sp1r28all_phase62a}). Generally, the number of possible observations that can resolve the disk with a signal-to-noise ratio greater than three increases with the total disk mass. In a very first approximation, the re-emission luminosity of a disk seen face-on depends linearly on the disk mass. In appendix \ref{a:mass} we summarize a more accurate, analytical prediction of the impact of the disk mass on the observability.\par
\begin{figure}[tp]
  \resizebox{\hsize}{!}{\includegraphics{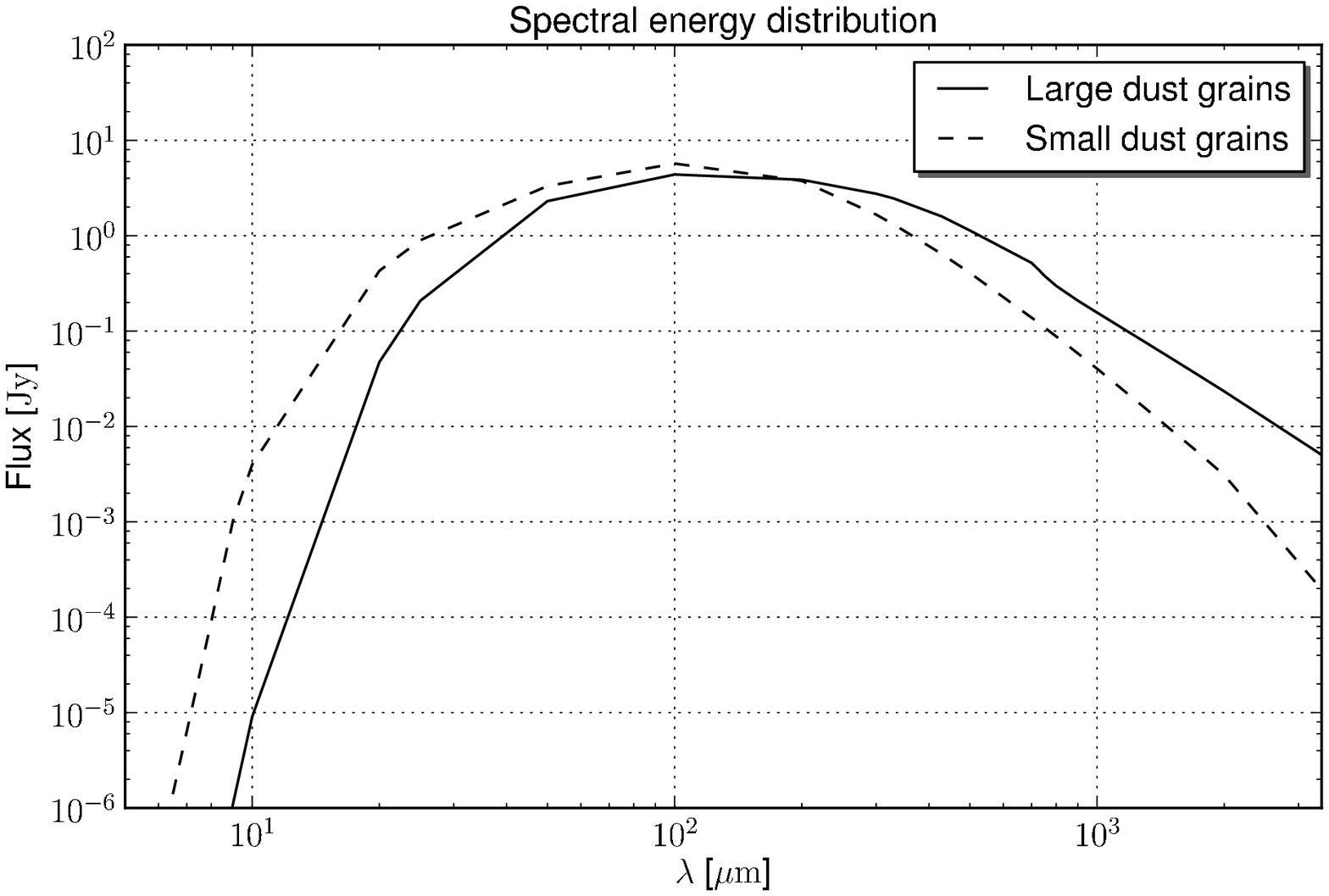}}
\caption{Differences in the SED resulting from different dust species. The solid line represents the large dust grains and the dashed line the small dust grains. In both cases an unperturbed circumstellar disk of $M = 2.67 \cdot 10^{-2} \, \rm M_\odot$ is used. A \textit{G type star} hosts this disk with an inner radius of $R_\textup{in} = 38 \, \rm AU$ and an outer radius of $R_\textup{out} = 171 \, \rm AU$. In the wavelength range between $730\, \rm \mu m$ and $3.3\, \rm mm$, the spectral index for the large grains is $\alpha_\text{spec, large} = 2.84 \pm 0.01$ and for the small dust grains $\alpha_\text{spec, small} = 3.96 \pm 0.01$. \label{fig:seds} }
\end{figure}

Besides the number of dust particles in the disk, which depends linearly on the total disk mass, the re-emission properties of the dust grains influence the disk luminosity. For this reason, we simulate the radiative transfer for two different dust species (see \S \, \ref{sec:radi}). Figure \ref{fig:qabs} shows the different absorption and re-emission properties of both dust populations. At a wavelength of $1\, \rm mm$, the re-emission ability of the small dust grains is just $1\%$ of the ability of the large dust grains. Differences are also reflected in the SEDs (see Fig. \ref{fig:seds}). While the maximum flux for both dust species is nearly the same, the wavelength of the maximum flux moves to shorter wavelengths for the smaller dust grains. Therefore a disk containing large dust particles re-emits more effectiv in the (sub)mm range, while a disk made up of small dust grains is more luminous in the infrared wavelength regime.This is also reflected by the spectral index:
\begin{equation}
 \alpha_\text{spec} = - \frac{\text{d}\log(F_\lambda)}{\text{d}\log(\lambda)}.
\end{equation}
In the wavelength range between $730\, \rm \mu m$ and $3.3\, \rm mm$ the spectral index for the large grains is $\alpha_\text{spec, large} = 2.84 \pm 0.01$ and for the small dust grains $\alpha_\text{spec, small} = 3.96 \pm 0.01$.
However, the flux of the disk in the case of smaller grain just drops to $10\%$ of the larger particle case. This is because the disk mass is the same in both scenarios. Owing to the average volume of the dust grains we need more smaller dust grains in comparison to the larger ones to produce the same disk mass. In our study, the effective dust grain radius of the small particles is about $10\%$ of the larger ones ($a_\textup{eff, small} \approx 0.1\, \cdot a_\textup{eff, large}$). Because of that, we need about thousand times more small dust grains for the same disk mass.\par
Although more dust in the disk influences the re-emitted flux of the disk, the re-emission properties of the particles are more important within our parameter space. The feasibility for a resolved detection drops to about a half when assuming smaller dust particles instead of large ones in every setup, independent of the underlying hydro- or magnetohydrodynamical simulation.\par

\subsection{Detectability of planet-induced Gaps \label{sec:resgaps}}
The individual parameters of our setup will influence the detectability of planet-induced gaps within the circumstellar disk. In the following we explore their impact for a hydrodynamical disk model with a planet with $\eta=0.001$. Since the quantity $\eta$ respresents the planet-to-star mass ratio (see \S \, \ref{sec:hydro}), the \textit{T Tauri star} ($M= 0.5 \, \rm M_\odot$) is orbited by a planet of a mass of $M = 0.5\, \rm M_\textup{Jup}$ and the \textit{Herbig Ae star} ($M= 2.5 \, \rm M_\odot$) is orbited by a planet of a mass of $M = 2.5\, \rm M_\textup{Jup}$.

\begin{figure}[tp]
  \resizebox{\hsize}{!}{\includegraphics{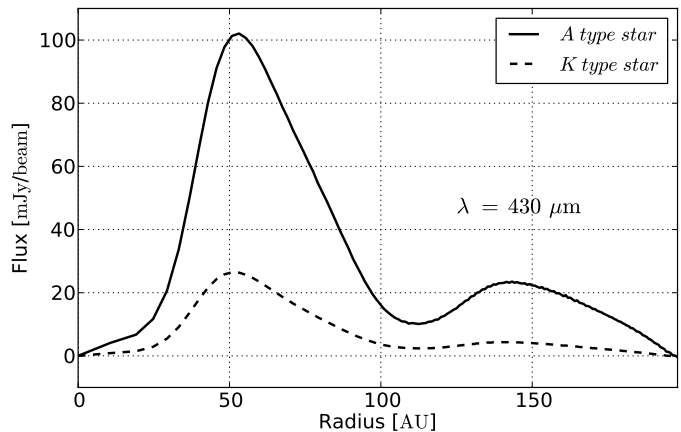}}
\caption{Comparison of the radial brightness distributions of identical disks around an {\textit{A type}} (solid line) and a {\textit{K type star}} (dashed line). \label{fig:AtoK}}
\end{figure}

\paragraph{Host star:} Without other heating sources, the stellar luminosity and temperature define the thermal structure and the re-emission of our disk models. Consequently, a luminous, high temperature star, like an \textit{A type star}, provides much better conditions for an optimal observation of a disk in re-emission radiation than a \textit{K type star} (for stellar parameters see \S \, \ref{sec:stars}). The \textit{A type star} engages a lot more energy in the disk and heats it up more efficiently. Because of the resulting higher dust temperature, the dust emission is enhanced. Therefore, the contrast between regions with and without dust increases. Figure \ref{fig:AtoK} shows an example for the resulting radial brightness distributions of a disk with a planet with $\eta=0.001$, which in the first case is illuminated by the \textit{A type star} and in the second by the \textit{K type star}. Through the increased disk remission flux in the first case, the gap is detectable as a minimum in the radial brightness distribution at $\approx 125 \, \rm AU$, while in the second case one only finds a saddle point at the location of the gap. In the parameter space considered in our study the  gap detection rate is higher by 60\% in the case of disks around \textit{A type stars}, than in the case of a \textit{K type star}.

\paragraph{ALMA configuration: \label{sec:config}} We consider a distance of $140 \, \rm pc$ for every simulated disk. Therefore, the required spatial resolution of an ALMA \-con\-figu\-ra\-tion only depends on the size of the structure that is intended to be observed. However, the sensitivity of ALMA decreases with increasing resolution. Accordingly, the optimal ALMA configuration is a compromise between the resolution and the sensitivity. Generally the best choice for observing a specific structure is to use an ALMA configuration with a synthesized beam that is slightly smaller than the structure itself. Thus, it is fully resolved, and a maximum flux is caught by the antennas. In general, for every combination of structure size and its re-emitted flux, there is an optimal ALMA configuration for the planned observation (see Fig. \ref{fig:sp1r28all_phase62a}). For example, the vast majority of observations at a wavelength of $430 \, \rm \mu m$ in the Fig. \ref{fig:sp1r28all_phase62b} is feasible with the 14$^{\rm th}$ ALMA configuration (max. baseline $\approx 1600\, \rm m$). This configuration is a valid compromise for a wide range of disk sizes.\par
\begin{figure}[t]
  \resizebox{\hsize}{!}{\includegraphics{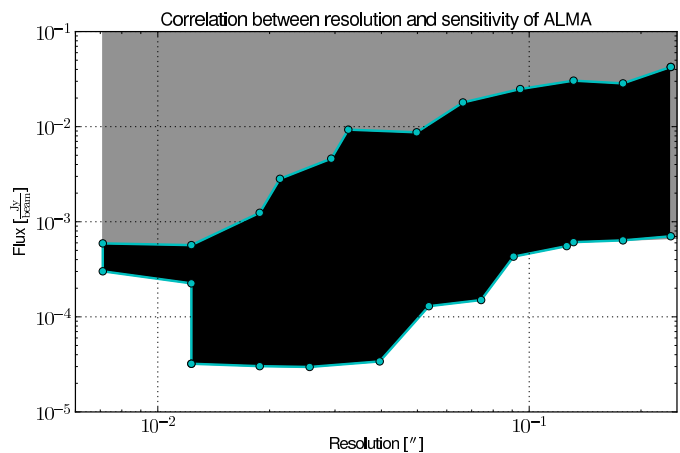}}
\caption{Overview of the correlation between resolution and sensitivity of ALMA. The resolution combines the observing wavelength and the ALMA array configuration. The dark area indicates $3\sigma$-detections within an exposure time of two hours. In the gray area observations are possible, but were not calculated in the current study. The ALMA detection limit is determined by the lower edge of the dark area. \label{fig:sensiall}}
\end{figure}

In this context, Fig. \ref{fig:sensiall} answers whether a structure is large and luminous enough to be observed by ALMA, or not.
Within the considered model space and combining the observing wavelength and ALMA array configuration to the resolution, a structure will be observable with ALMA if its combination of required luminosity and required angular resolution is located in the dark area in the middle of Fig. \ref{fig:sensiall}. The lower limit of this dark area is the detection limit of ALMA. Above, in the gray area, observations are possible, but these and the ones represented through the regions on the left and right of the dark area are not covered by our parameter space.\par
In our setup ALMA allows for detections of circumstellar disks down to a scale of $\approx 2\, \rm AU$ (Fig. \ref{fig:sensiall} for a distance of $140 \, \rm pc$). A finer resolution is limited by the sensitivity and position-dependent maximum exposure time. (Below an elevation of $20^{\circ}$ over the horizon an observation should not be carried out, \citealtads{casa}.)
\begin{figure}[t]
  \resizebox{\hsize}{!}{\includegraphics{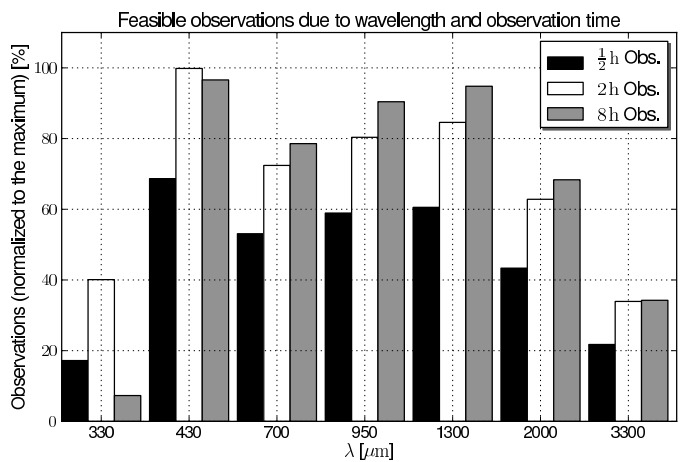}}
\caption{Overview of the influence of the exposure time ($\frac{1}{2} \, \rm h$: black bar, $2 \, \rm h$: transparent bar, $8 \, \rm h$: gray bar) and wavelength on the feasibility of detecting a gap in the disk. The underlying model contains all simulations (all disk masses, sizes, and stellar types) for a hydrodynamical simulation (SP1) with a planet ($\eta = 0.001$) and large dust grains. The observing probability is normalized to the maximum at $430\, \rm \mu m$ and an exposure time of $2\, \rm h$. \label{fig:wellel}}
\end{figure}

\paragraph{Observing wavelength and exposure time: \label{sec:wavel}} The impact of wavelength and exposure time is illustrated in Fig. \ref{fig:wellel}. This histogram shows that it is most likely to observe a gap in a disk in the wavelength range from $430 \, \rm \mu m$ to $2000 \, \rm \mu m$ for each simulated exposure time (total wavelength range in our setup: $330 \, \rm \mu m$ to $3300 \, \rm \mu m$). While the resolution of ALMA is higher in the lower part of this regime, the density structure of the disk is better traced at wavelengths longer than $950\, \mu \rm m$, because of the low optical depth of the disk (see appendix \ref{a:mass}), the low temperature ($\approx 20\, \rm K$) of the most parts of the disk, and the resulting high re-emission ability in the mm-wavelength regime.
In the context of our model setup, the wavelengths $430 \, \rm \mu m$ and $1300 \, \rm \mu m$ under the respective atmospheric conditions are good choices for searching gaps in circumstellar disks, because a maximum number of observations is possible (Fig. \ref{fig:wellel}). At both wavelengths, ALMA detects gaps down to a disk mass of $2.67 \cdot 10^{-4} \, \rm M_\odot$ (see Fig. \ref{fig:sp1r28all_phase62a}, \textit{bottom} and \ref{fig:sp1r28all_phase62b}, \textit{middle}).\par In comparison to the wavelength range from $430 \, \rm \mu m$ to $2000 \, \rm \mu m$, the possibility of detecting a gap within a circumstellar disk decreases strongly at a wavelength of $330 \, \rm \mu m$. Here, gaps can be traced only in the most massive disks around the most luminous stars (see Fig. \ref{fig:sp1r28all_phase62b}, bottom). Even with optimum observation conditions the thermal noise dissipates the incoming signal too much. This effect is worse the nearer the object is located to the horizon. At wavelengths longer than $2000 \, \rm \mu m$, the re-emitted flux of the disk decreases below the detection limit of ALMA (see \S \, \ref{sec:config}) or the resolution of the interferometer is no longer sufficient.\par
The exposure time (see Fig. \ref{fig:wellel}) affects the signal-to-noise ratio and the coverage of the u-v plane in Fourier space. With the exception of the shorter wavelengths in our setup, the signal-to-noise ratio grows, as expected, with the square root of the exposure time. The decrease in the likelihood of tracing a gap at the shorter wavelength between a two-hour and an eight-hour observation is again a result of the atmosphere. During an eight-hour observation our object is located just $\approx 20^\circ$ over the horizon for about one hour. Therefore, the signal-to-noise ratio decreases rapidly.\par
In about 66\% of the simulated observations, in which a gap was detected in a two-hour observation, this gap can already be traced within half an hour (see Fig. \ref{fig:wellel}). An even shorter exposure time carries the risk of too low u-v coverage for the more extended configurations. On the other hand, an eight-hour observation will not produce much better results than a two-hour one.

\paragraph{Planet-disk interaction:}
\begin{figure}[t]
  \resizebox{\hsize}{!}{\includegraphics{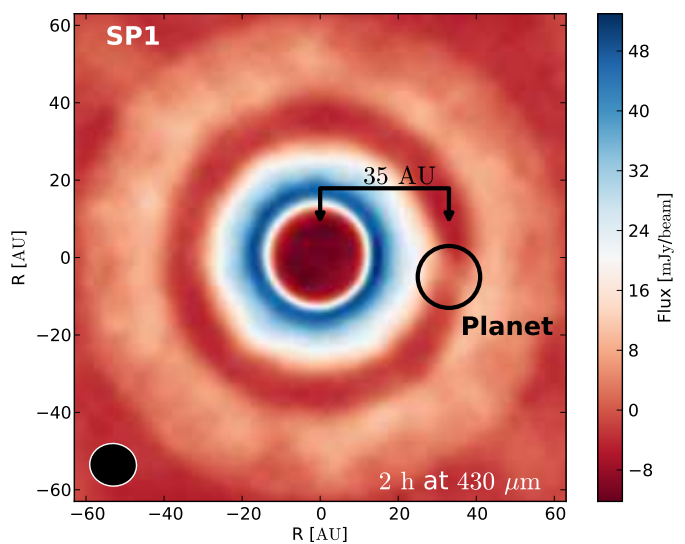}}
\caption{Example of a simulated observational result. Around a Herbig Ae star even the circumplanetary region can be detected. The simulated observation of the disk with $M = 2.67 \cdot 10^{-2} \, \rm  M_\odot$ is made at wavelength of $430 \, \rm \mu m $ with the 18$^{\rm th}$ ALMA configuration. The exposure time amounts to two hours. Dust phase: large dust grains. The inner radius of the disk is $R_\textup{in} = 14 \, \rm AU$, the semi-major axis of the planet $R_\textup{pl} = 35 \, \rm AU$, and an outer radius of $R_\textup{out} = 63 \, \rm AU$.\label{fig:observed}}
\end{figure}
Besides the gap, the planet-disk interaction causes structures like spiral waves and circumplanetary density accumulation \citepads{1984ApJ...285..818P}. 
The spiral waves created by the Lindblad resonances \citepads{1958StoAn..20....4L} pass through the entire disk. They are clearly visible in the face-on density profiles shown in Fig. \ref{fig:f1}. However, the (sub)mm brightness contrast between the spiral arms and the background disk surface is too low to be indicated by ALMA at a distance of $140\, \rm pc$. But spiral arms resulting from self-gravitation of massive, compact disks can potentially be observed by ALMA, because of the much higher densities in this kind of disk \citep{2010MNRAS.407..181C}.\par 

The circumplanetary disk is not as large in size as the gap or the spiral arms. However, this high density area in the low density gap produces a high level of contrast \citepads{2005ApJ...619.1114W}. Therefore, it can be observed by ALMA even with neglected planetary heating (Fig. \ref{fig:observed}). The planet as an additional heating source, in fact, has to be considered in a realistic model and thus the contrast level will even increase. The strength of the planet-induced disk structures depends on the planet-to-star mass ratio $\eta$ (see \S \, \ref{sec:hydro}). Besides $\eta = 0.001$, we investigate the detectability of gaps for hydrodynamical simulations with $\eta = 0.0001$. However, in this case planet-disk interaction can not be traced by ALMA.\par 
Generally, sufficiently massive planets perturb a circumstellar disk massively. However, a planet cannot be detected unambiguously in the SED. This has already been investigated by \citetads{2005ApJ...619.1114W}. Our results confirm this statement.

\begin{figure*}[htbp]
\centering
\includegraphics*[width=15cm]{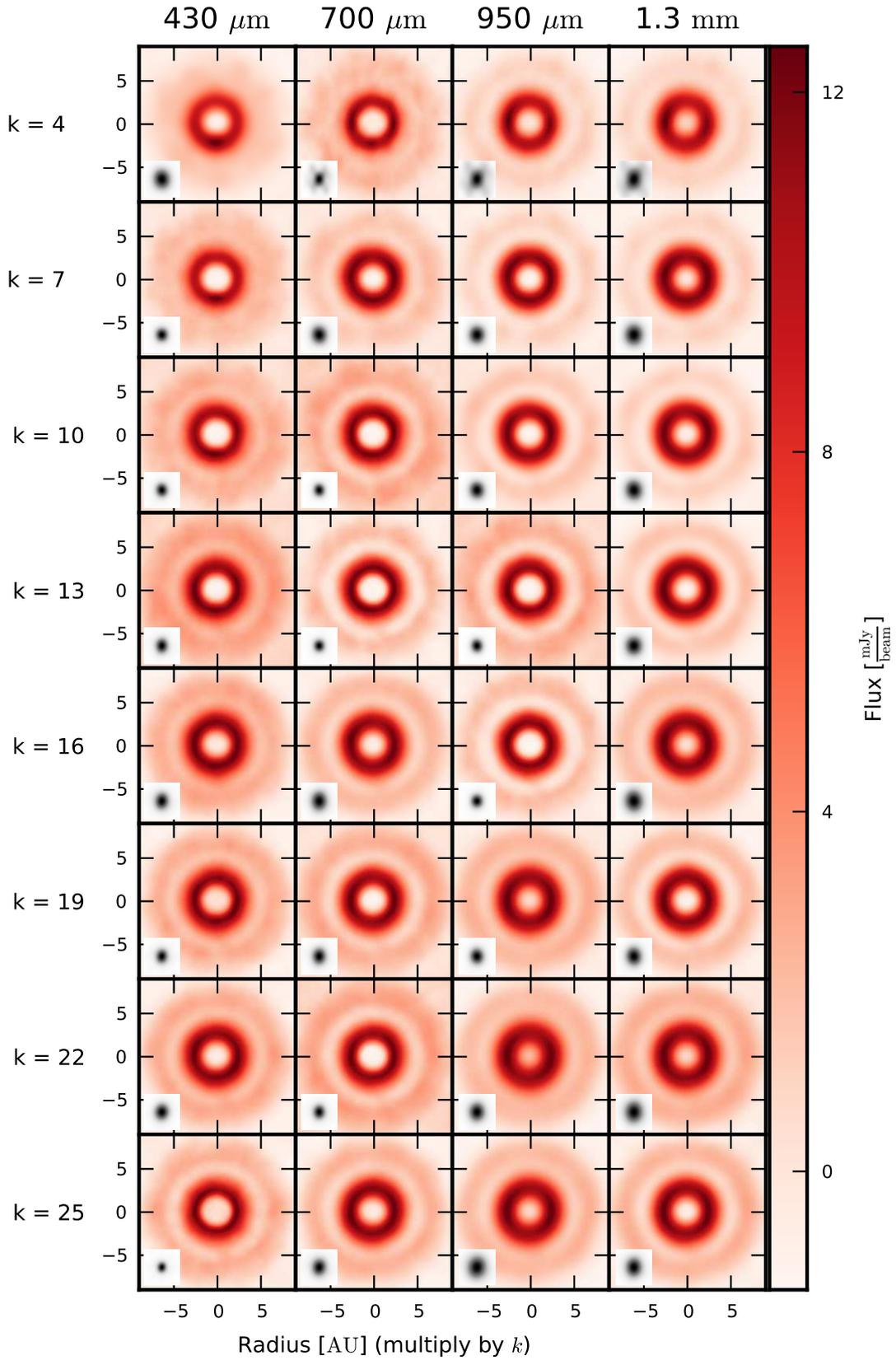}
\caption{Selected examples for the detection of gap (illustrated as a bright ring at a radius of $k\, \times \, 5 \, \rm AU$) in circumstellar disks, which corresponds to the Butterfly Star disk model. The shown scale of the disk size has to be multiplied by the scaling factor $k$, shown for every line on the left.  \label{fig:butter}}
\end{figure*}
\paragraph{Optimal parameters for an observation:} Based on our investigations, we propose to use an exposure time of about two hours, a wavelength of $430 \, \rm \mu m$ or $1300 \, \rm \mu m$, and the 14$^{\rm th}$ ALMA configuration (max. baseline $\approx 1600\, \rm m$) for the search of planet-induced structures in circumstellar disks in the Taurus-Auriga star formation region.

\subsection{Application for real objects \label{sec:butter}}
In the following we briefly diskuss the possibility of tracing planet-induced structures in a disk that is comparable to the specific, well-studied case of the Butterfly Star. The Butterfly Star is a perfectly edge-on seen Class I source in Taurus \citepads{2003ApJ...588..373W}. For this purpose we use the \textit{mass} of cut-outs from the Butterfly Star disk model (see appendix \ref{a:butter}) and  \textit{disk density structure} from a hydrodynamical disk model with a planet with $\eta=0.001$. In this case the disk size already defines the disk mass. The masses that we consider are shown in Table \ref{tbl:massen}. The disk size still is a function of the scaling factor $k$ (see \S \, \ref{sec:model}).
Our results are shown in Fig. \ref{fig:butter}. With the exception of the smallest disks in our setup ($R_\textup{out} = 9 \, \rm AU$ corresponding to the scaling factor $k = 1$), gaps are traced for every disk size ($k = 1$ is not plotted in Fig. \ref{fig:butter}). The synthesized beam is about the same size in every image and always comparable to the corresponding gap size. Thus, ALMA can trace gaps with nearly all semi-major axis at several wavelengths with identical resolution. Besides the actual gap detection therefore statements on the dust in the disk are an additional outcome of an observation.\par
We find that ALMA will be able to indicate a planet-induced gap in a disk at a distance of $140\, \rm pc$ if the planet has a semi-major axis of $10 \, \rm AU$ (diameter of the gap $\approx 2\, \rm AU$). \citetads{2005ApJ...619.1114W} detected this structure, which was caused by a giant planet with a semi-major axis of $ 5 \, \rm AU$ at a wavelength of $330 \, \rm \mu m$ at a distance of $100 \, \rm pc$. 
The main differences are the wavelength of the observation and the boundary conditions of the hydrodynamical simulations.
However, for observations at a wavelength of $330 \, \rm \mu m$, meaningful results are uncertain because of the absorption properties of the atmosphere (\S \, \ref{sec:wavel}). In general, it is possible to trace planet-induced structures in $140\, \rm pc$ distant circumstellar disks at AU-scale with a wavelength of about $430 \, \rm \mu m$ and a small amount of material between star and planet, because of the resulting higher contrast between the gap and its outer edge.

\begin{table}[t]
\caption{Disk dimensions and masses for the real object application. \label{tbl:massen}}
\centering
\begin{tabular}[c]{l|ccc|r}
\hline \hline
$k$ & R$_\textup{in}$& R$_\textup{planet}$& R$_\textup{out}$ & Mass \\
    & {[$\rm AU$]} & {[$\rm AU$]} & {[$\rm AU$]} &  {[$\rm M_\odot$]} \\ 
\hline
1 & 2 & 5 & 9 & $0.0021$ \\
4 & 8 & 20 & 36 & $0.0076$\\
7 & 14 & 35 & 63 & $0.0127$\\
10 & 20 & 50 & 90 & $0.0176$\\
13 & 26 & 65 & 117 & $0.0215$\\
16 & 32 & 80 & 144 & $0.0271$\\
19 & 38 & 95 & 171 & $0.0318$\\
22 & 44 & 110 & 198 & $0.0364$\\
25 & 50 & 125 & 225 &$0.0410$\\
\end{tabular} 
 \tablefoot{Dimension and masses of the cut-outs from the Butterfly Star disk model ($M_\textup{tot} = 0.07 \, \rm M_\odot$). Besides the scaling factor $k$, the inner and outer disk radii (R$_\textup{in}$ and R$_\textup{out}$) and the planet major semi-axis (R$_\textup{planet}$) are listed.}
\end{table}

\subsection{Effects of magnetohydrodynamical turbulence \label{sec:mhd}}

\begin{figure}[t]
  \resizebox{\hsize}{!}{\includegraphics{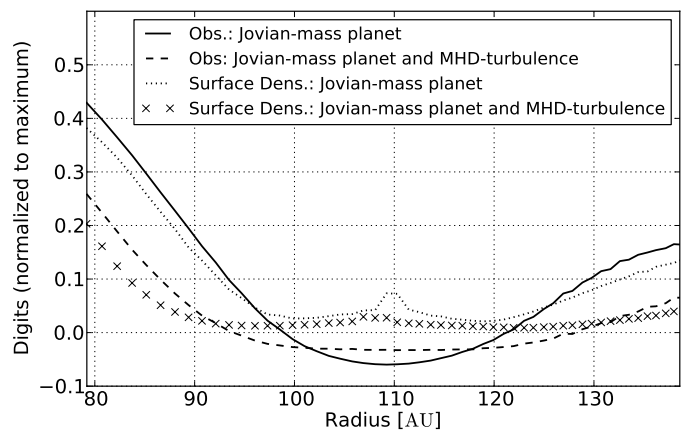}}
\caption{Measuring the gap size in the radial brightness distribution resulting from an ALMA observation and in the surface density distribution. The solid line represents the radial brightness profile and the dotted line the surface density of a disk model hosting a planet with $\eta=0.001$ without turbulence (SP1). The dashed line represents the radial brightness profil and the crosses the surface density of correspoding disk model with magnetohydrodynamical turbulence (SPT6). The central star is a Herbig Ae.  \label{fig:sp1-spt6}}
\end{figure}

\begin{figure}
  \resizebox{\hsize}{!}{\includegraphics{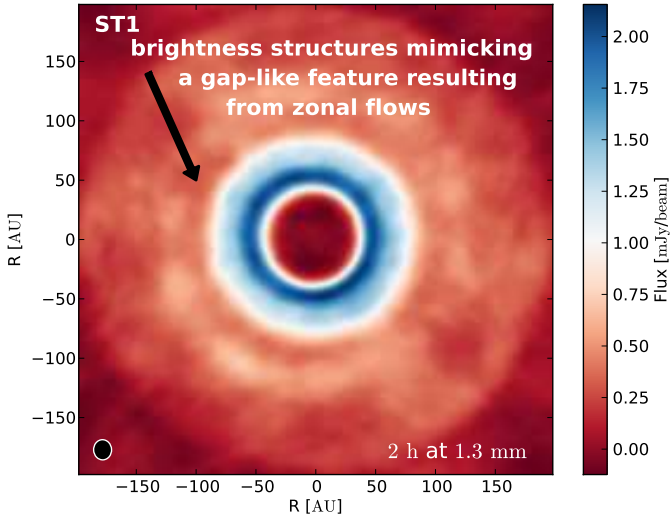}}
\caption{Observability of brightness structures mimicking gap like features resulting from zonal flows. The disks is made of large dust grains around a Herbig Ae star at a wavelength of $1300 \, \rm \mu m$ with the 20$^{\rm th}$ ALMA configuration. ($M_\textup{disk} = 2.67 \cdot 10^{-2} \, \rm M_\odot$) \label{fig:turbu}}
\end{figure}

Our setup includes equivalent magnetohydrodynamical simulations for every pure hydrodynamical simulation. Their classification and diskussion is given in \citetads{2011ApJ...736...85U} in detail. First we focus on a comparison between the magnetohydrodynamical simulation and the pure hydrodynamical simulation of a disk with a planet with $\eta = 0.001$. \citetads{2011ApJ...736...85U} have already investigated that the presence of magnetohydrodynamical turbulence in the disk enlarges the gap size (see also \citealtads{2003ApJ...589..543W}; \citealtads{2003MNRAS.339..993N}). ALMA again allows us to measure the size of the gap if the synthesized beam of the configuration is about half of the gap size. By this, a different gap size is not the result of an under-resolved observation. For example, Fig. \ref{fig:sp1-spt6} compares the radial brightness distribution (normalized to its maximum) of the pure hydrodynamical simulation and the magnetohydrodynamical simulation and the corresponding surface density structures (also normalized to their maxima) in an identical disk setup. The observation measures the gap size in good agreement with the surface density profiles. The differences are mainly caused by the scale of the synthesized beam and the signal-to-noise ratio of the observation.\par
The magnetohydrodynamical turbulence also has an impact in the case of simulations of planets with $\eta=0.0001$. While gaps cannot be traced in the pure hydrodynamical case, it will be feasible to detect gaps if magnetic fields are considered. It is possible to detect this gap in about 10\% of the cases in which a gap induced by a planet with $\eta=0.001$ has already been traced.\par 
However, a larger gap is not a proof of the importance of turbulence and dissipative effects in the disk. It also can be formed by a planet containing more mass (see \S \, \ref{sec:intro}). Therefore, further observations that can determine the planet mass are needed to analyze the origin of the gap size, i.e. a measurement of the orbital period and the semi-major axis of the planet. In general, the presence of magnetohydrodynamical turbulence improves the detectability of gaps. The observability of other planet-induced, large-scale structures is equivalent to the already diskussed pure hydrodynamical case. Thus, the results of \S \, \ref{sec:resgaps} also apply to the magnetohydrodynamical case.\par

\begin{figure}[tp]
  \resizebox{\hsize}{!}{\includegraphics{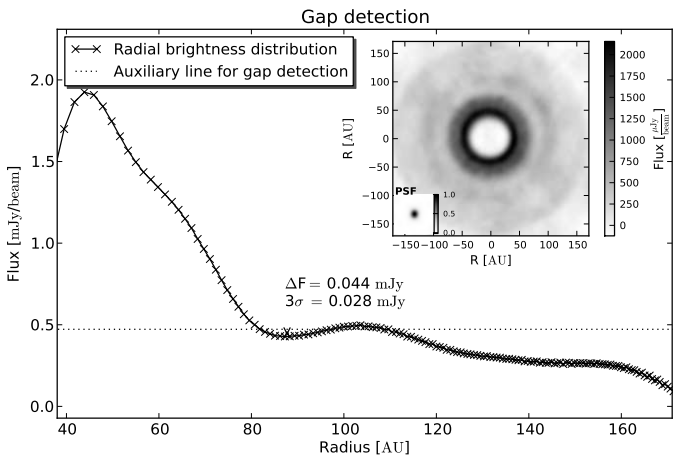}}
\caption{Radial brightness profile (solid line) corresponding to Fig. \ref{fig:turbu}. The auxiliary line is given dotted.  \label{fig:f18}}
\end{figure}

Additionally, we explore the observability of zonal flows (see \S \, \ref{sec:hydro}) in circumstellar disks. In very few cases ($\approx 20$ images) a gap-like structure resulting from zonal flows is detectable  at a wavelength of $1300 \, \rm \mu m$ with the ALMA configurations 18 to 22. In particular a massive and extended disk out of large dust particles around a bright stars are needed for a detection. Figure \ref{fig:turbu} illustrates an example of these results. The corresponding radial brightness profile is shown in Fig. \ref{fig:f18}. Zonal flows have an impact on the dust distribution in the disk and can trigger the formation of planetesimals (see \S \, \ref{sec:hydro}). To distinguish between gaps that are induced by planets and gap-like features resulting from zonal flows, further multi wavelength observations of the object are necessary.

\section{Conclusion}
 \label{sec:conclusion}

We have performed a large-scale parameter study of the observability of planet-induced structures in circumstellar disks in the (sub)mm wavelength regime with ALMA.
The basic assumptions of the investigated parameter space are typical T Tauri objects at a distance of $140\,\rm pc$ and a declination of $\approx 20^\circ$.
Based on this, we explored the impact on the observation properties of six central stars, nine disk sizes with outer radii from $9 \, \rm AU$ to  $225 \, \rm AU$,
15 total disk masses in the range between $2.67 \cdot 10^{-7} \, \rm M_\odot$ and $4.10 \cdot 10^{-2} \, \rm M_\odot$, different observation wavelengths $\left( \left[330\, \mu \text{m}, 3 \, \text{mm} \right] \right)$ and exposure times ($\frac{1}{2} \, \rm h$, $2  \, \rm h$, $ 8 \, \rm h $), 14 ALMA configurations, two dust grain size distributions, and disk turbulence due to magnetic fields. For nearly all conceivable parameter combinations large-scale disk structures are detected by ALMA. Even without considering the planet as an additional heating source, the circumplanetary high density area is traced. Gaps are clearly identified for gap width that are larger than $2\, \rm AU$. A  more detail resolution of the disk structures is limited by the sensitivity and the object's position-dependent maximum exposure time. Nevertheless, ALMA allows measuring gap sizes and therefore distinguishing between planet-to-star ratios or underlying disk models, which both influence the gap size. Through the observing wavelength range of ALMA, in particular, large dust particles enhance the feasibility of tracing planet-induced disk structures. For typical protoplanetary disks, the disk mass affects the observability only slightly. In general, a wavelength of $430 \, \rm \mu m$ or $1300\, \rm \mu m$, in combination with the 14$^{\rm th}$ ALMA configuration (max. baseline $\approx 1600\, \rm m$) and an exposure time of about two hours, is ideal for future surveys aimed at detecting disk structures, in particular those induced by planets. In magnetohydrodynamical disk models, it is more likely to trace planet-induced gaps, because the presence of a magnetic field enlarges the gap size \citepads{2011ApJ...736...85U}. We find that zonal flows are detectable with ALMA. A major fraction of the presented simulations and more overview charts are available online at \url{http://www1.astrophysik.uni-kiel.de/~placid}.\par
To enhance the understanding of the gap and planet formation processes and to further improve the knowledge of extrasolar planetary systems, a deep link of the radiative transfer to the hydro- and magnetohydrodynamical simulations is necessary. The radiation of the star affects the gap form and enlarges the vertical extension especially of the outer edge \citepads{2012ApJ...749..153J}, which finally influences the detectability.
Another goal is to implement the accurately modeled emission of the planet into the simulation setup.

\begin{acknowledgements}
We acknowledge financial support by the German Research Foundation (J.P. Ruge: WO 857/10-1; H.H. Klahr: KL 14699-1). Computations were partially performed on the Bluegene/P supercomputer of MPG and the THEO cluster of MPIA at the Rechenzentrum Garching (RZG) of the Max Planck Society and the support of the International Max Planck Research School (IMPRS) of Heidelberg. The authors would like to acknowledge the helpful comments of the anonymous referee.
\end{acknowledgements}

\appendix

\section{Atacama Large (Sub)Millimeter Array}
\label{a:alma}

\begin{figure}[t]
  \resizebox{\hsize}{!}{\includegraphics{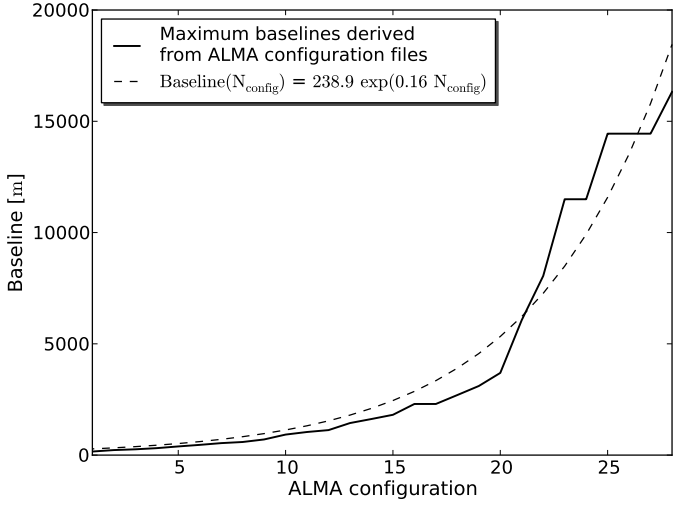}}
\caption{Relationship between ALMA configuration and maximum baseline of the array. Maximum baselines derived from the CASA ALMA configuration files are shown as solid line. An exponential fit onto this data is shown as dashed line. Only the maximum baseline is shown, since the antennas within this maximum baseline are arranged in different configurations, the resolution of ALMA can also change. \label{fig:baseline}}
\end{figure}

We used the CASA 3.2 simulator \citepads{2012ASPC..461..849P} to predict the observability of the selected star/planet/disk configurations. This simulation tool comes with 28 ALMA array configuration files containig the position of every antenna in this configuration. In Fig. \ref{fig:baseline} the relationship between the maximum array baseline and the corresponding ALMA configuration is plotted. However, not only the maximum baseline, but also the u-v coverage determines the quality of the reconstructed maps \citepads[i.e.][]{2012A&A...539A..89B}.
For further information about ALMA configurations and CASA see \url{http://almascience.eso.org/}.

\section{Deriving the model disk mass from the mass of the Butterfly Star disk}
\label{a:butter}
We use the results \citetads{2003ApJ...588..373W} applied to the modeling of the disk around the Butterfly Star to define the required disk mass in our model. The parameters derived for the disk model of this source are comparable to those for other circumstellar disks ($M_\textup{total} = 0.07 \, \rm M_\odot$, see \citealtads{2012A&A...543A..81M}; \citealtads{2011A&A...533A..89G}; \citealtads{2009A&A...505.1167S}; \citealtads{2008A&A...478..779S}; \citealtads{2008ApJ...674L.101W}; \citealtads{2003ApJ...588..373W}).
\citetads{2003ApJ...588..373W} used a disk density profile of the form
\begin{equation}
 \rho_\textup{disk}\left(r,z\right) = \rho_0 \left(\frac{R_{\star}}{r}\right)^{\alpha} \exp\left\{-\frac{1}{2} \left[ \frac{z}{h\left(r\right)} \right]^2 \right\}, \label{glg:shakura}
\end{equation}
where $R_{\star}$ is the stellar radius and $h\left(r\right)$ is the scale height of the disk:
\begin{equation}
 h\left(r\right) = h_0 \left(\frac{r}{R_{\star}}\right)^{\beta}\textup{.} \label{glg:hr}
\end{equation}
From the volume integration of this disk density profile adapted to the Butterfly Star disk ($\alpha = 2.367 $, $\beta = 1.29 $), we derive the total disk mass $M(R)$ within the radius $R$:
\begin{equation}
 M(R) = 3.70 \cdot 10^{-4}\, \textup{M}_\odot \cdot \left(R \left[ \textup{AU} \right] \right)^{0.923} \label{glg:buttermass}.
\end{equation}
If we interpret one of our simulated and scaled disks as a cut-out from the Butterfly Star disk, we can use the correlation Eq. (\ref{glg:buttermass}) to estimate the total disk mass of such a cut-out. 
Since the outer radius in our studies is a function of the scaling factor $k$ (see \S \, \ref{sec:model}), there are nine resulting disk masses, already listed in Table \ref{tbl:massen}.

\section{Optical depth and total disk mass}
 \label{a:mass}

\begin{figure}[htbp]
  \resizebox{\hsize}{!}{\includegraphics{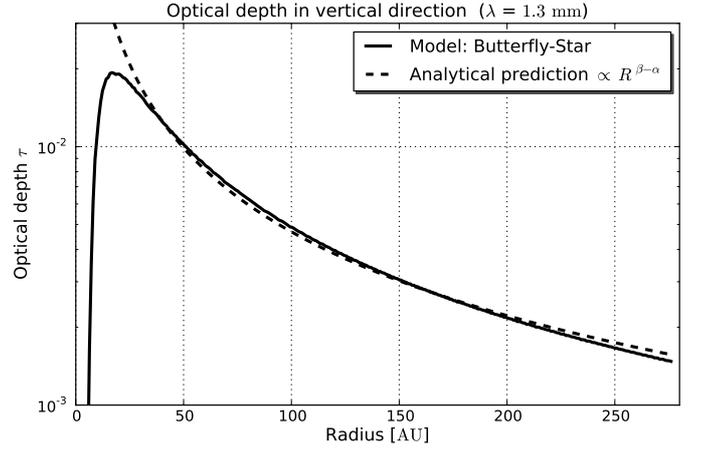}}
\caption{Comparison between the analytical (dashed line) and numerical model (solid line) for the optical depth in vertical direction in a circumstellar disk corresponding to the model for the Butterfly Star as diskussed in appendix \ref{a:mass}. The disk consists of large dust grains.\label{fig:tau}}
\end{figure}
\begin{figure}[htbp]
  \resizebox{\hsize}{!}{\includegraphics{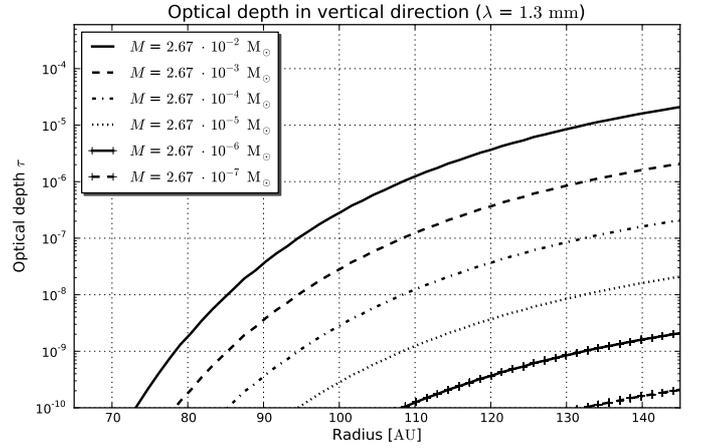}}
\caption{Impact of the total disk mass on the optical depth in vertical direction of an unperturbed disk with an inner radius of $R_\textup{in} = 38 \, \rm AU$ and an outer radius of $R_\textup{out} = 171 \, \rm AU$ at a wavelength of $1.3 \, \rm mm$. The disk consists of large dust grains. Different line styles represent different total disk masses. Please see the legend in the upper right of the figure.\label{fig:tau_vert_butter}} 
\end{figure}

The total disk mass is one of the main parameters for the observability. Especially the highest disk masses feature good conditions for an observation (Figs. \ref{fig:sp1r28all_phase62a} and \ref{fig:sp1r28all_phase62b}). The luminosity of an optically thin debris disks linearly depends on the dust mass \citepads{2012A&A...544A..61E}, but for optically thick disks, like protoplanetary disks, this estimation is more difficult.
For the optical thickness $\tau_\nu$ along a way $\textup{d}z$ in the vertical direction of the disk with a mass density of the dust $\rho_\textup{dust}$ and a disk density profile $\rho_\textup{disk}$, we achieve
\begin{eqnarray}
\tau_\nu &\approx& \int\limits_{-\infty}^{+\infty}\pi \cdot Q_\textup{ext}(\nu) \cdot a_\textup{eff}^2 \cdot  \frac{\rho_\textup{disk}}{\rho_\textup{dust} \cdot \left(\frac{4}{3} \pi \, a_\textup{eff}^3 \right)}\, \textup{d}z \\
	&=& \frac{3 \, Q_\textup{ext}(\nu)}{4 \, a_\textup{eff} \, \rho_\textup{dust}} \, \int\limits_{-\infty}^{+\infty} \rho_\textup{disk}\, \textup{d}z = \frac{3 \, Q_\textup{ext}(\nu)}{4 \, a_\textup{eff} \, \rho_\textup{dust}} \Sigma.
\end{eqnarray}
Here $\Sigma$ is the surface density of the disk and $Q_\textup{ext}$ the extinction efficiency.
Based on this analytical estimation we test a numerical implementation in MC3D for the disk around the Butterfly Star (see Fig. \ref{fig:tau}). For the density structures from the hydro- and magnetohydrodynamical simulations (see \S \, \ref{sec:hydro}), we find that the optical thickness of the disk in the vertical direction is too low to influence the radiative transfer in the Rayleigh-Jeans wavelength regime (see Fig. \ref{fig:tau_vert_butter}).

With a constant disk size, dust grain size, and dust mass density, the only process for increasing the mass within the disk is to increase the number of particles (linearly dependence). While the disk is optically thin in the vertical direction (see Fig. \ref{fig:tau_vert_butter}), the effective luminous surface of the disk $A$ depends on the number of dust particles within the model space, so we assume
\begin{equation}
 A = A_0 \cdot \frac{M}{M_0} \label{glg:A}.
\end{equation}

\begin{table}[t]
\caption{Values and standard deviation for the fitted $\gamma$-parameter from Eq. \ref{glg:T}. }
\label{tbl:fit}
\begin{center}
\begin{tabular}{lc}
\hline\hline
total disk mass & $\gamma$ \\
\hline
$ 2.67 \cdot 10^{-2} \, \rm M_\odot$ & $ 0.64 \pm  3.96 \cdot 10^{-5}$ \\
$ 2.67 \cdot 10^{-3} \, \rm M_\odot$ & $ 0.58 \pm  3.56 \cdot 10^{-5}$ \\
$ 2.67 \cdot 10^{-4} \, \rm M_\odot$ & $ 0.49 \pm  1.02 \cdot 10^{-5}$ \\
$ 2.67 \cdot 10^{-5} \, \rm M_\odot$ & $ 0.43 \pm  1.53 \cdot 10^{-7}$ \\
$ 2.67 \cdot 10^{-6} \, \rm M_\odot$ & $ 0.41 \pm  8.63 \cdot 10^{-8}$ \\
$ 2.67 \cdot 10^{-7} \, \rm M_\odot$ & $ 0.42 \pm  1.33 \cdot 10^{-7}$ \\
\hline
\end{tabular}
\end{center}
 \end{table}

\begin{figure}[t]\centering
  \resizebox{\hsize}{!}{\includegraphics*{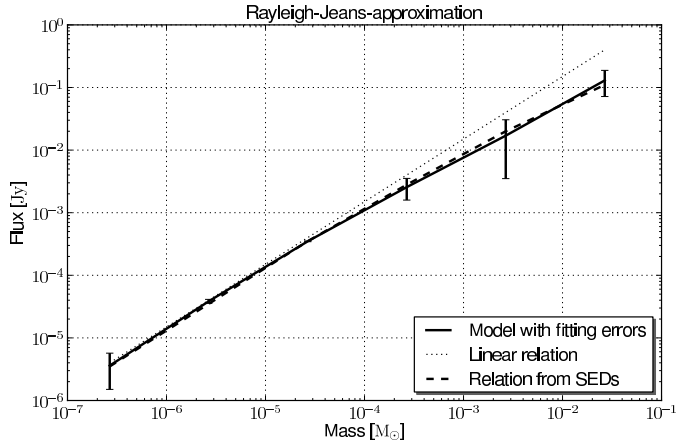}}
\caption{Influence of the total disk mass on the re-emitted flux of an unperturbed circumstellar disk. A comparison between a linear model (dotted line) that is useful for very low mass disks, a model on the basis of the Rayleigh-Jeans-law (solid line), and the flux simulated in the radiative transfer (dashed line) is shown. Every disk has an inner radius of $R_\textup{in} =  38 \, \rm AU$ and an outer radius of $R_\textup{out} =  171 \, \rm AU$, and the hosting star is a \textit{G type star} (\S \, \ref{sec:stars}). The disk consists of large dust grains. \label{fig:rayleigh}}
\end{figure}

This fact is sufficient for debris disks, but for a circumstellar disk the optical thickness along a line of sight from the star has to be considered, too. Besides the linear increasing effective luminous surface of the disk, we find that the temperature profile of the disk midplane can be used to approximate the re-emitted flux in the Rayleigh-Jeans-regime. We assume the following temperature profile:
\begin{equation}
 T(R) = C_0\, R^{- \gamma}. \label{glg:T}
\end{equation}
We take $C_0 =  300\, \rm K$ and $\gamma \in \left[0,1\right]$ ($\gamma$ depends on the dust mass $M$).
Using the Rayleigh-Jeans-law:
\begin{equation}
 B_\lambda(T) \approx \frac{2\, \pi c \, k_\textup{B}\, T}{\lambda^4} \label{glg:rayleigh}
\end{equation}
and the gray-body approximation of the dust, we get the flux $F$ of a radial symmetric disk seen face-on by inserting Eq. (\ref{glg:T}) into Eq. (\ref{glg:rayleigh}) and integrating over the radius coordinate:
\begin{equation}
 F  = \frac{2\, \pi^2 c \, k_\textup{B}}{\lambda^4}\, Q_\textup{abs}(\lambda) \cdot C_0 \, A \int\limits_{R_\textup{in}}^{R_\textup{out}} R^{- \gamma} \, \textup{d}R. \label{glg:F}
\end{equation}
Neglecting the case $\gamma = 1$, we can estimate the flux $F$:
\begin{equation}
F = \frac{2\, \pi^2 c \, k_\textup{B}}{\lambda^4 \, M_0}\, Q_\textup{abs}(\lambda) \cdot A_0 \cdot C_0 \cdot \frac{\left(R_\textup{out}^{1 - \gamma} - R_\textup{in}^{1 - \gamma} \right)}{1 - \gamma}  \cdot M.
\end{equation}
This result is comparable to \citetads{2000prpl.conf..533B}, who used a constant disk temperature.
For testing our model, we use six circumstellar disks of the same size but different disk masses. From the SEDs we get their flux at a wavelength of $ 1300 \, \rm \mu m$. Thereafter we fit the parameter $\gamma$ from Eq.(\ref{glg:T}) to the corresponding temperature profiles. The results are shown in Fig. \ref{fig:rayleigh}, and the fitting values of  $\gamma$ are listed to the left of the figure. The benefit of the method is the easily prediction of the disk temperature profile and its mass from a spatially resolved millimeter observation.

\bibliography{ruge2012}

\end{document}